\documentclass[reprint,amsmath,amssymb,aps,superscriptaddress,dvipsnames]{revtex4-2}

\usepackage{amsmath,amssymb,amsfonts}
\usepackage{algorithmic}
\usepackage{graphicx}
\usepackage{textcomp}

\usepackage{graphicx}
\usepackage{dcolumn}
\usepackage{bm}

\usepackage[utf8]{inputenc}
\usepackage[T1]{fontenc}
\DeclareMathAlphabet{\mathcal}{OMS}{cmsy}{m}{n}

\usepackage{braket}
\usepackage{amsmath}
\usepackage{caption}
\usepackage{subcaption}
\usepackage{gensymb}
\usepackage{graphicx}
\usepackage{mwe}
\usepackage{stackengine}
\usepackage{float}
\usepackage{physics}
\usepackage{etoolbox}

\usepackage{qcircuit}

\usepackage[dvipsnames]{xcolor}
\usepackage{hyperref}
\hypersetup{
  breaklinks=true,
  colorlinks=true,
  allcolors=BlueViolet,
}
\usepackage[capitalise]{cleveref}

\usepackage[normalem]{ulem}

\newcommand{\zz}{\mathcal{ZZ}}

\newcommand{\Yb}{\textsuperscript{171}\rm{Yb}\textsuperscript{+}}

\newcommand{\appsection}[1]{\section{\MakeUppercase{#1}}}

\def\framezero/{\texttt{frame0}}
\def\frameone/{\texttt{frame1}}
\def\toneone/{\texttt{tone1}}
\def\tonezero/{\texttt{tone0}}

\def\BibTeX{{\rm B\kern-.05em{\sc i\kern-.025em b}\kern-.08em
    T\kern-.1667em\lower.7ex\hbox{E}\kern-.125emX}}

\begin{document}
\newtoggle{ieeeformat}
\togglefalse{ieeeformat}

\iftoggle{ieeeformat}{

\history{Date of publication xxxx 00, 0000, date of current version \today}
\doi{10.1109/TQE.2020.DOI}
	

\title{Realization and Calibration of Continuously Parameterized Two-Qubit Gates on a Trapped-Ion Quantum Processor}
\author{
\uppercase{Christopher G. Yale}\authorrefmark{1,*},
\uppercase{Ashlyn D. Burch}\authorrefmark{1,*,$\dagger$},
\uppercase{Matthew N. H. Chow}\authorrefmark{1,2,3,*, $\ddagger$},
\uppercase{Brandon P. Ruzic}\authorrefmark{1},
\uppercase{Daniel S. Lobser}\authorrefmark{1},
\uppercase{Brian K. McFarland}\authorrefmark{1},
\uppercase{Melissa C. Revelle}\authorrefmark{1},
\uppercase{Susan M. Clark}\authorrefmark{1}}
\address[1]{Sandia National Laboratories, Albuquerque, New Mexico 87123, USA}
\address[2]{Department of Physics and Astronomy, University of New Mexico, Albuquerque, New Mexico 87131, USA}
\address[3]{Center for Quantum Information and Control, CQuIC, University of New Mexico, Albuquerque, New Mexico 87131, USA}
\address[*]{These authors contributed equally}
\address[$\dagger$]{Present address: Oak Ridge National Laboratory, Oak Ridge, Tennessee 37831, USA}
\address[$\ddagger$]{Present address: HRL Laboratory, LLC, Malibu, California 90265, USA}
\tfootnote{This research was supported by the U.S. Department of Energy, Office of Science, Office of Advanced Scientific Computing Research Quantum Testbed Program. Sandia National Laboratories is a multi-mission laboratory managed and operated by National Technology \& Engineering Solutions of Sandia, LLC (NTESS), a wholly owned subsidiary of Honeywell International Inc., for the U.S. Department of Energy’s National Nuclear Security Administration (DOE/NNSA) under contract DE-NA0003525. This written work is authored by an employee of NTESS. The employee, not NTESS, owns the right, title and interest in and to the written work and is responsible for its contents. Any subjective views or opinions that might be expressed in the written work do not necessarily represent the views of the U.S. Government. The publisher acknowledges that the U.S. Government retains a non-exclusive, paid-up, irrevocable, world-wide license to publish or reproduce the published form of this written work or allow others to do so, for U.S. Government purposes. The DOE will provide public access to results of federally sponsored research in accordance with the DOE Public Access Plan.}

\markboth
{Yale \headeretal: Realization and Calibration of Continuously Parameterized Two-Qubit Gates on a Trapped-Ion Quantum Processor}
{Yale \headeretal: Realization and Calibration of Continuously Parameterized Two-Qubit Gates on a Trapped-Ion Quantum Processor}

\corresp{Corresponding author: Christopher G. Yale (email: cgyale@sandia.gov).}}{

\preprint{AIP/123-QED}
\title[]{\color{black}Realization and Calibration of Continuously Parameterized Two-Qubit Gates on a Trapped-Ion Quantum Processor}

\author{Christopher G. Yale}
\thanks{These authors contributed equally}
\email{cgyale@sandia.gov}
\affiliation{Sandia National Laboratories, Albuquerque, New Mexico 87123}

\author{Ashlyn D. Burch}
\thanks{These authors contributed equally; Present address: Oak Ridge National Laboratory, Oak Ridge, Tennessee 37831, USA}
\affiliation{Sandia National Laboratories, Albuquerque, New Mexico 87123}

\author{Matthew N. H. Chow}
\thanks{These authors contributed equally; Present address: HRL Laboratory, LLC, Malibu, California 90265, USA}
\affiliation{Sandia National Laboratories, Albuquerque, New Mexico 87123}
\affiliation{Department of Physics and Astronomy, University of New Mexico, Albuquerque, New Mexico 87131, USA}
\affiliation{Center for Quantum Information and Control, CQuIC, University of New Mexico, Albuquerque, New Mexico 87131, USA}

\author{Brandon P. Ruzic}
\affiliation{Sandia National Laboratories, Albuquerque, New Mexico 87123}

\author{Daniel S. Lobser}
\affiliation{Sandia National Laboratories, Albuquerque, New Mexico 87123}

\author{Brian K. McFarland}
\affiliation{Sandia National Laboratories, Albuquerque, New Mexico 87123}

\author{Melissa C. Revelle}
\affiliation{Sandia National Laboratories, Albuquerque, New Mexico 87123}

\author{Susan M. Clark}
\affiliation{Sandia National Laboratories, Albuquerque, New Mexico 87123}

\date{\today}}

\begin{abstract}

Continuously parameterized two-qubit gates are a key feature of state-of-the-art trapped-ion quantum processors as they have favorable error scalings and show distinct improvements in circuit performance over more restricted maximally entangling gatesets. In this work, we provide a comprehensive and pedagogical discussion on how to practically implement these continuously parameterized  M\o{}lmer-S\o{}rensen gates on the Quantum Scientific Computing Open User Testbed (QSCOUT), a low-level trapped-ion processor. To generate the arbitrary entangling angles, $\theta$, we simply scale the amplitude of light used to generate the entanglement. However, doing so requires careful consideration of amplifier saturation as well as the variable light shifts that result. As such, we describe a method to calibrate and cancel the dominant fourth-order effects, followed by a dynamic virtual phase advance during the gate to cancel any residual light shifts, and find a linear scaling between $\theta$ and the residual light shift. Once, we have considered and calibrated these effects, we demonstrate performance improvement with decreasing $\theta$. Finally, we describe nuances of hardware control to transform the XX-type interaction of the arbitrary-angle M\o{}lmer-S\o{}rensen gate into a phase-agnostic and crosstalk-mitigating ZZ interaction.
\end{abstract}

\iftoggle{ieeeformat}{

\titlepgskip=-15pt}{}

\maketitle

\section{Introduction}
\label{sec:introduction}

Trapped-ion quantum computers have grown in size, complexity, and performance metrics over the past decade, with flagship systems now exhibiting two-qubit gate fidelities greater than 0.998 and register sizes in excess of 30 qubits~\cite{moses2023, ionqms}. One of the workhorses of these processors is the two-qubit entangling M\o{}lmer-S\o{}rensen (MS) gate, proposed in 1999~\cite{Molmer1999, sorenson1999}, and experimentally demonstrated in 2003~\cite{Leibfried2003}. It has now become the most prevalent entangling operation available in many commercial trapped-ion quantum processors~\cite{nam2020, baldwin2020zz, Loschnauer2024, aqt-website} as well as academic and national laboratory efforts~\cite{egan2021,erhard2021lq,clark2021}. The two-qubit MS gate finds application in a wide array of complex quantum algorithms for simulation and optimization protocols, but also has a potential role in quantum networking applications~\cite{Inlek2017}, error correction schemes~\cite{Heussen2024}, quantum metrology~\cite{Broz2023}, and quantum machine learning~\cite{Alam2022}. While work has been done to improve the effective implementation of this gate in experiment~\cite{ballance2016, Azuma_2024, Kang2023, Webb2018, ruzic2024}, both stochastic and coherent errors still remain limitations to gate fidelity, and therefore it is vital to understand how to best calibrate against error sources that limit qubit coherence and gate operation times.

Here, we provide a pedagogical discussion of the calibration of a two-qubit arbitrary-angle MS gate. In particular, we discuss pulse shaping and motional mode selection, the approach used to generate arbitrary amounts of entanglement, and the impact and mitigations for light shifts that arise during the interaction. The MS gate is based on a transverse Ising-type interaction, where the unitary for the gate may be written:
\begin{equation}
    U_{MS}(\theta, \phi) = \exp{-i\frac{\theta}{2}\left(\hat{\sigma}_{\phi, i} \otimes \hat{\sigma}_{\phi, j}\right)}
\end{equation}
where $\theta$ is the rotation angle, $i$ and $j$ are ion-label indices, and $\hat{\sigma}_{\phi} = \cos(\phi)\hat{\sigma}_x + \sin(\phi)\hat{\sigma}_y$ is a rotation about the equatorial Bloch-sphere axis defined by phase $\phi$, where $\hat{\sigma}_x$ and $\hat{\sigma}_y$ are the usual Pauli operators. The typical implementation of the gate is a maximally entangling Clifford operation occurring at a $\theta = \frac{\pi}{2}$ rotation, i.e. $MS(\pi/2)$.

In many noisy intermediate-scale quantum, or NISQ, computing applications, generating arbitrary amounts of entanglement is favorable as gate error tends to decrease with decreasing gate angle~\cite{nam2020, moses2023}. These variable-angle gates, $MS(\theta)$ where $\theta \in [0,\pi/2]$, are particularly useful for decomposing arbitrary unitary operations via Cartan's KAK decomposition~\cite{tucci2005}. This decomposition means any arbitrary two-qubit $SU(4)$ unitary can either be broken down into three $MS(\pi/2)$ or broken into three $MS(\theta)$~\cite{baldwin2022qv, campbell2023, yale2024superstaq}. Clear performance improvements have been demonstrated when compiling circuits via an arbitrary-angle entangling gateset rather than a more limited gateset with only maximally entangling gates~\cite{yale2024superstaq}.

\begin{figure}[h]
\centering
\includegraphics[width=0.47\textwidth]{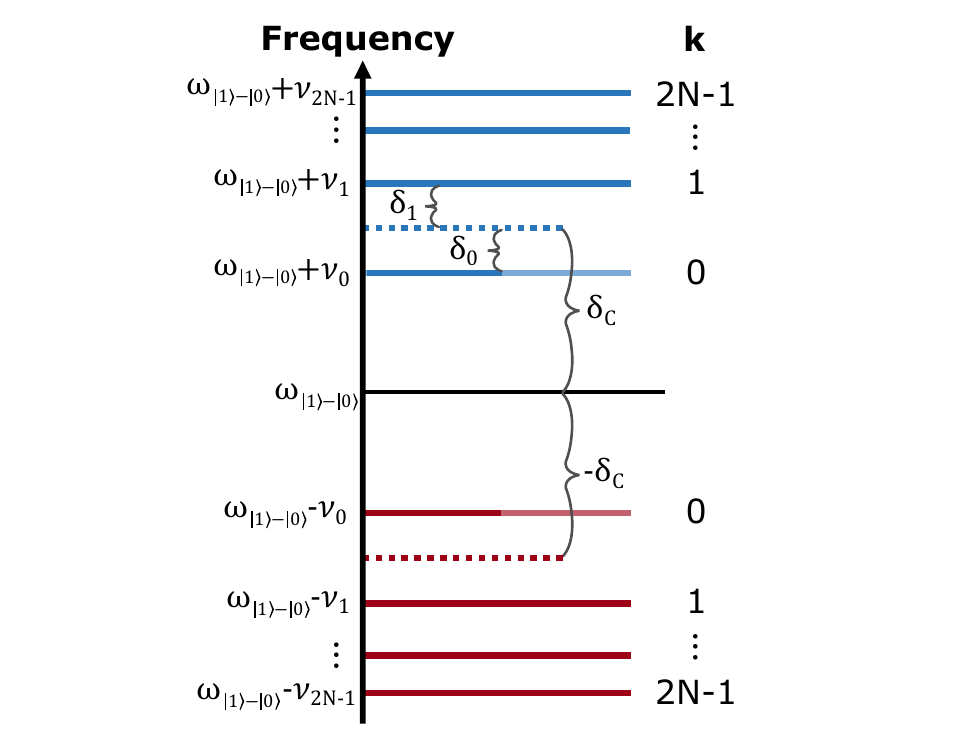}
\caption{Illustration of sideband transitions and Raman detunings used in an $MS(\theta)$ gate. The hyperfine states $\ket{0}$ and $\ket{1}$ are separated with an energy splitting of $\omega_{\ket{1}-\ket{0}} = 12.643$ GHz. For a chain of N ions, there are also 2N radial motional modes $k$ where $\{k:0 ... 2N-1\}$, and each blue (red) mode is characterized by frequency $\nu_{k} (-\nu_{k})$. The Raman tones applied to the IA beams are symmetrically detuned from the blue (red) motional mode $k$ by $\delta_{k} (-\delta_{k})$. We also denote $\delta_c (-\delta_c)$ indicating the equivalent blue (red) detunings from the carrier transition such that $\delta_k = \delta_c - \nu_k$.}
\label{fig:detunings}
\end{figure}

The experiments detailed here are performed on the Quantum Scientific Computing Open User Testbed (QSCOUT) located at Sandia National Laboratories~\cite{clark2021}. The QSCOUT register consists of a linear chain of $\Yb$ ions whose hyperfine states $^{2}S_{1/2} \ket{F=0, m_F = 0} (\ket{0})$ and $^{2}S_{1/2} \ket{F=1, m_F = 0} (\ket{1})$ serve as the qubit states, split by 12.6 GHz. We utilize Raman transitions to drive between the qubit levels for all laser-based gates. The single-photon detuning of the Raman transition is positioned between the $^{2}P_{1/2}$ and $^{2}P_{3/2}$ levels, roughly 33 THz away from $^{2}P_{1/2}$.  During an $MS(\theta)$ gate, motional sidebands of the two target qubits are addressed by red- and blue-detuned Raman transitions to implement the necessary spin-dependent force for the entangling interaction~\cite{Molmer1999, sorenson1999}. As these Raman transitions must be sensitive to the motion of the ions, they are performed in a counter-propagating configuration -- achieved by applying a single-toned, all-encompassing beam from one direction orthogonal to the full length of the chain (i.e. the `global' beam) and individual addressing (IA) beams on each of the target ions from the opposite side. Each of the IA beams consist of two tones, which are symmetrically blue- and red-detuned by $\delta_c$ from the qubit splitting (i.e. carrier transition) as shown in Fig.~\ref{fig:detunings}. To control these tones we use our custom control hardware system Octet, an RF system-on-chip (RFSoC) device that is specifically designed to support two tones per channel. Each Octet channel controls an acousto-optic modulator (AOM) for the global and IA beams, and provides independent amplitude, phase, and frequency modulation specified either as splines or discrete changes. More details of the beam geometry and Octet hardware control can be found in Ref. ~\cite{clark2021}. 

In support of the discussion that follows we introduce the Hamiltonian for the bare MS gate interaction:

\begin{equation}
\label{eq:ms-hamiltonian}
H(t) = -\frac{i\hbar}{2}\sum_{i,k}{\sigma_{x,i}\eta_{k,i}\Omega_{i} a_ke^{-i\delta_kt} + h.c.}
\end{equation}

Here, $k$ denotes any of the 2N motional modes along the radial directions within a chain of N ions (i.e. two orthogonal radial principal axes each host a manifold of N motional modes), $\eta_{k,i}$ is the Lamb-Dicke parameter of mode $k$ for ion $i$, $\Omega_{i}$ is the Rabi drive frequency targeting ion $i$, and $a^{\dagger}_k$ and $a_k$ are the creation and annihilation operators of mode $k$ characterized by frequency $\frac{\nu_k}{2\pi}$. As shown in Fig. \ref{fig:detunings}, $\delta_k$ is the detuning of the applied light from the blue or red sideband resonance frequency of mode $k$.

\section{Frequency Robustness via Gaussian-Pulse Shape and Mode Choice} \label{sec:freq_robust}

Motional modes are subject to frequency drift, and so frequency and amplitude modulation can be utilized as a tool to combat these drifts and other sources of noise when the driving field becomes entangled with the ion’s motional modes~\cite{Leung2018, Kang2021}. We use amplitude modulation along with a specific fixed detuning to limit these types of errors. As described in Ref.~\cite{ruzic2024}, we find that using a spectrally compact pulse shape can minimize displacement errors that result in residual spin-motion entanglement after a gate. This pulse shape requirement is fulfilled by a Gaussian envelope which is approximated as a spline and applied to the amplitude of the RF waveform driving the AOMs.

We choose the fixed frequency such that the detuning balances contributions to $\theta$ from multiple modes. From this, the gate becomes significantly more robust to variations in motional frequency up to 10 kHz~\cite{ruzic2024}. This implementation is robust to symmetric changes in mode frequencies typical of variations in driving RF power, but not to antisymmetric changes due to DC voltage or stray field instability. We choose the detuning for each pair of ions such that both of the nearest detuned modes,$k_-, k_+$, have strong $\eta_{k,i}$ for each of the ions $i$ in the pair and have signs such that the contributions of the nearest modes add constructively. More expressly, we select modes such that $|\eta_{i,k_+}\eta_{j,k_+} - \eta_{i,k_-}\eta_{j,k_-}|$ is maximized.

For even numbers of ions, this mode selection criteria is straightforward within a single radial mode manifold. However, for odd chain lengths, the center ion does not participate in antisymmetric modes where the behavior on either side of the center ion is equal and opposite. Therefore, we select operating detunings slightly nearer to a symmetric mode for gates involving the center ion and forgo the robustness to motional frequency changes provided by motional mode balancing.

\section{Realization of Arbitrary-Angle Gates} \label{sec:arbangle}
With our waveform and mode choice determinations, we now turn our attention to the ability to generate arbitrary amounts of entanglement. 
From the Hamiltonian shown in Eq.~\ref{eq:ms-hamiltonian}, it can be seen that choice of laser intensity, laser detuning $\delta_k$, and gate duration (which respectively determine the Rabi drive frequency $\Omega_{i}$, the interaction strength, and interaction time $\tau$) can be used separately or together to set $\theta$.  In this work, we apply an amplitude scaling factor to the laser intensity to vary $\theta$. Specifically, we apply the scaling factor to the global beam AOM while keeping the IA beam pulse shape and size fixed. We note this approach is similar to Ref. \cite{nam2020}, in which they applied an overall scale factor to the intensity of the applied laser light and found a roughly linear relationship between gate error and gate angle. Alternatively, Ref.~\cite{moses2023} describes an approach in which detuning and gate duration were varied together for gates with $\theta>0.075\pi$, whereas for $\theta<0.075\pi$, only laser intensity was varied. 
It should be noted that a similar roughly linear relationship was found between gate error and angle despite the different methodologies. 

\subsection{Saturation Response of Acousto-Optic Modulator}
To make full use of the available optical power, for $MS(\pi/2)$, we drive the global beam AOM such that the peak of the Gaussian pulse shape may require an applied RF amplitude that approaches the peak diffraction efficiency of the AOM. Thus, it can introduce distortion as the AOM has a nonlinear saturation response. However, we calibrate this response and apply the appropriate correction. 

The optical response of the AOM to an RF drive is given by \cite{AAOptoelectronicsNote}:
\begin{equation}
    I_1 \propto \sin^{2}\left(\frac{\pi}{2}\frac{a}{a_{\rm sat}}\right),
    \label{eq:aomsat-power}
    \end{equation}
where $I_1$ is the intensity of the first-order diffracted beam, $a$ is a unitless software scaling value for the RF amplitude, and $a_{\rm sat}$ is the corresponding scaled RF amplitude at which the AOM response saturates.
As $\Omega$ is proportional to the square root of the intensity, we write down the equation with proportionality constant $\Xi$: 
    \begin{equation}
        \Omega(a) = \Xi \sin \left(\frac{\pi}{2}\frac{a}{a_{\rm sat}}\right).\label{eq:aom-omega}
    \end{equation}
    
\begin{figure}
    \centering
    \includegraphics[width=0.47\textwidth]{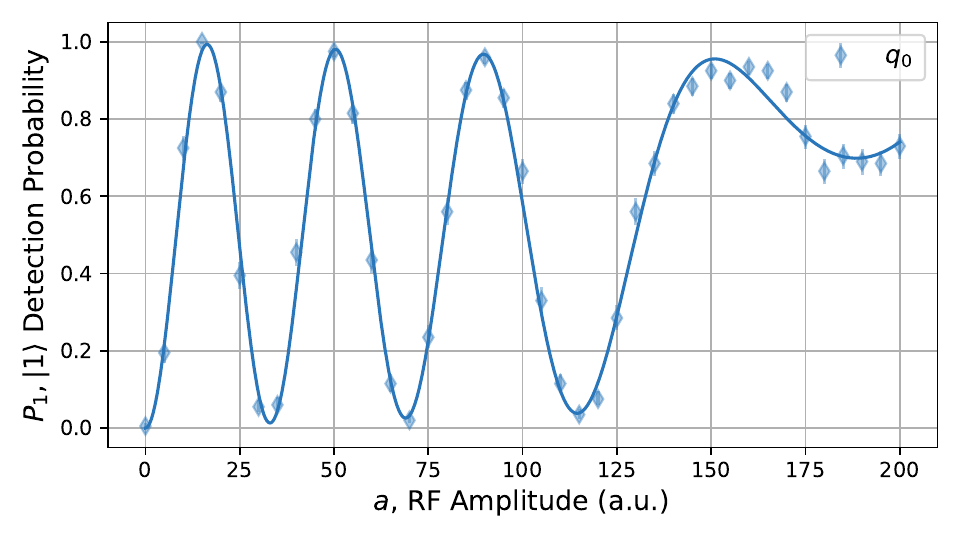}
    \caption{The RF amplitude ($a$) applied to the global beam AOM is varied and the resulting Rabi oscillation fit to Eq.~\ref{eq:ampscan-fitfn} in order to determine saturation parameters. Uncertainty interval shown is a $1\sigma$ Wilson score and are roughly the size of the points.
    }
    \label{fig:ampscan}
\end{figure}

To calibrate $a_{\rm sat}$ and $\Xi$, we initialize in the $\ket{0}$ state and perform a Rabi oscillation with fixed $t$ while sweeping $a$. The resulting data is fit to
    \begin{align}
        P_1 &= \frac{1}{2}\left(1 - \exp(-\Omega(a) t/\xi)\cos(\Omega(a) t)\right) 
        \label{eq:ampscan-fitfn}
    \end{align}
where $P_1$ is the probability of detecting the ion in state $\ket{1}$, $\xi$ is an exponential decay constant to account for dephasing of the Rabi oscillation. Example data for this fit using a gate duration of $t=50\,\mu$s is shown in Fig.~\ref{fig:ampscan}, where we find $a_{\rm sat} = 188.5(6)$ and $\Xi = 2\pi \times 73.6(1)\,$kHz. Uncertainty on these parameters is the square root of the diagonal elements of the covariance matrix from the fit.

The $a_{\rm sat}$ parameter is a device-specific parameter related to the RF response of each specific AOM. For the global beam AOM, we find it to be relatively stable for a given frequency, so we recalibrate it infrequently. The $\Xi$ parameter, on the other hand, is related to the optical power in both legs of the Raman transition, and the optical powers of each beam are recalibrated frequently due to gradual beam misalignment and optical degradation. 

\subsection{Sequential Applications of $MS(\theta)$}

\begin{figure}[h!!!!!]
    \centering
    \includegraphics[width=0.47\textwidth]{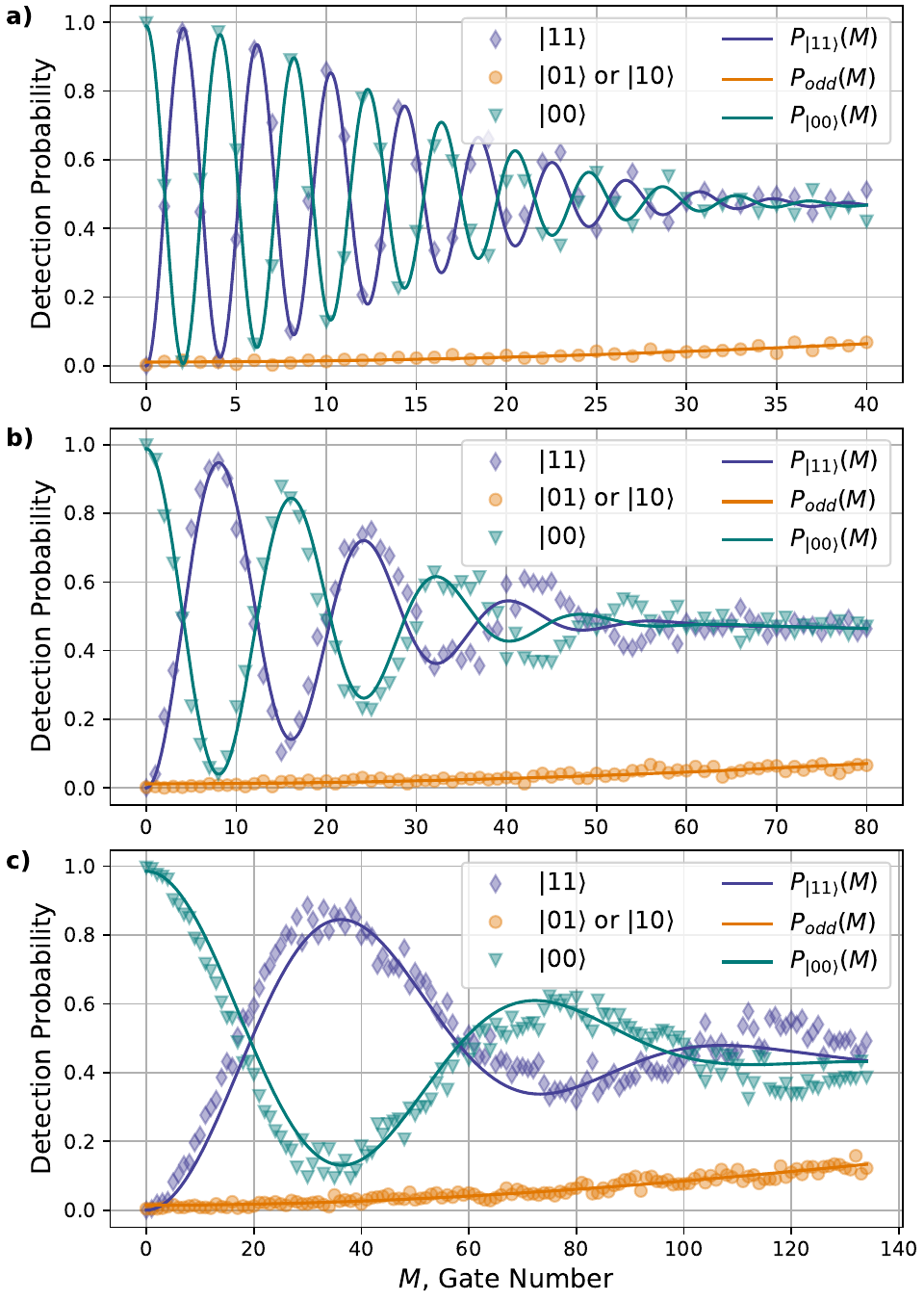}
    \caption{Repeated applications of the MS gate show less dephasing per gate for small $\theta$. Ideal performance would show oscillation between $\ket{00}$ and $\ket{11}$ in steps of $\theta = \frac{\pi}{2}, \frac{\pi}{8},$ and $\frac{\pi}{32}$ for variable gate number (M) for the a), b) and c), respectively.  }
    \label{fig:allmsloops}
\end{figure}

To confirm our scaling methodology is accurate, we perform repeated applications of the gate at various angles ($\theta = \pi/2, \pi/8,$ and $\pi/32$) and observe the population transfer between $\ket{00}$ and $\ket{11}$. We note that the gates described here are the fully calibrated versions including the light shift cancellation described in Section~\ref{sec:lightshifts}. From these oscillatory population transfer measurements, we can estimate the scaling of gate error vs $\theta$. As shown in Fig.~\ref{fig:allmsloops}, we prepare in the $\ket{00}$ state and drive oscillations between the even-parity states ($\ket{00}$ and $\ket{11}$), with leakage to odd-parity states ($\ket{01}$ and $\ket{10}$). As discussed in~\cite{Benhelm2008}, the amplitude oscillations of the even parity states and overall gate fidelity will follow a Gaussian decay when looping over multiple gates as there can be random shifts in the applied laser frequency.

We model the probability of detecting even-parity states $P_{\rm even}$ according to 
\begin{equation}
    \label{eq:p_even}
    P_{\rm even}(M) = 1 - P_{\rm odd}(M)
\end{equation}
in which $P_{\rm odd}(M)$ reflects detection of leakage into the odd-parity states modeled as
\begin{equation}
    \label{eq:oddparitydecay}
    P_{\rm odd}(M) = \frac{1}{2}\left(1 - A\exp(-\frac{M^2}{2 M^2_{\sigma, \rm odd}})\right)
\end{equation}
where $M$ is the number of $MS(\theta)$ gates applied, $A$ is a fit parameter that allows for a finite SPAM offset, and $M_{\sigma, \rm odd}$ is the Gaussian standard deviation representing leakage rate into the odd-parity subspace.

With this, we can model the dephasing of the Rabi oscillation within the even-parity subspace with an additional Gaussian decay envelope:
\begin{equation}
    \label{eq:Rabi11}
    \begin{split}
    P_{\ket{11}}(M, \theta)& = \frac{P_{\rm even}(M)}{2}\\
   &\times\left(1-\exp(-\frac{M^2}{2M^2_{\sigma, \rm even}})\cos(\theta M)\right)
    \end{split}
\end{equation}
and
\begin{equation}
    \label{eq:Rabi00}
    \begin{split}
    P_{\ket{00}}(M, \theta) &= \frac{P_{\rm even}(M)}{2}\\
    &\times\left(1+\exp(-\frac{M^2}{2M^2_{\sigma, \rm even}})\cos(\theta M)\right)
    \end{split}
\end{equation}

where $M_{\sigma, \rm even}$ is the standard deviation of the Gaussian decay envelope.

We first fit the measured values of $P_{\rm odd}$ according to Eq.~\ref{eq:oddparitydecay}, then fit $P_{\ket{11}}$ to Eq.~\ref{eq:Rabi11}. For $\theta = \frac{\pi}{2}, \frac{\pi}{8}, $ and $\frac{\pi}{32}$ we find $M_{\sigma, \rm odd} = $ 83(3), 157(4), and 179(2) gates and $M_{\sigma, \rm even} = $ 12.9(5), 19.8(8), and 48 (1) gates, respectively. 

We note that these decay envelopes do not fully replicate the actual behavior of the system. In particular at high numbers of repeated gates, the Gaussian decay overestimates the decay of the even parity states. Likewise, there is no clear relation between the odd decay profile and the even decay profiles. This suggests this phenomenological model is not fully representative of all of the potential limiters: decoherence, dephasing, amplitude fluctuations, heating, and axial motion. Regardless, there is a trend for increasing coherence with decreasing $\theta$.

\section{Light Shifts and Frame Rotations} \label{sec:lightshifts}

Due to the significant laser powers required to achieve these $MS(\theta)$ gates, it is imperative that we are cognizant of the light shifts that result. Light shifts induce additional energy splittings on the qubit levels as a result of the Autler-Townes, or AC Stark, effect.  In particular, these additional energy splittings on the qubit levels will affect the precession of the qubit relative to the bare rotating frame leading to phase errors in the gate if uncompensated. In this section, we discuss the sources of our light shifts during the $MS(\theta)$ gate and our methods for cancellation. 

\subsection{Origin of Light Shifts}

The typical, second-order, light shift is largely minimized by our choice of wavelength for the Raman laser~\cite{Campbell2010}.  However, we use a frequency comb to generate Raman transitions, and thus a significant contribution from fourth-order light shifts remains~\cite{lee2016lightshifts}. These fourth-order light shifts may be understood as detuned two-photon Raman coupling. This coupling occurs both between the usual qubit states ($\ket{F=0, m_F=0}$ and $\ket{F=1, m_F=0}$) and due to state coupling from $\ket{F=0, m_F=0}$ to the $\ket{F=1, m_F = \pm 1}$ Zeeman sublevels in the $^{2}S_{1/2}$ hyperfine manifold. The dominant contribution to fourth-order light shifts during the MS gate in our system is from the detuned two-photon drive of the qubit carrier transition, and thus, we will only consider these here. As the effective Raman Rabi rate is approximately $\Omega_{\rm eff} \approx 2\pi\times 125$\,kHz and the radial motional mode frequencies range from 2 to 2.5 MHz, fourth-order shifts of $|\Omega_{\rm eff}^2/4\delta| \approx 2\pi \times 2$\,kHz are expected from the rf tones driving the red and blue sideband. Since the sign of $\delta$ is opposite for the red and blue drives, if $\Omega_{\rm eff}$ can be made equal for the two transitions during the MS gate, these contributions cancel, as is the case for a continuous-wave laser~\cite{Kirchmair2009}.

\begin{figure}[h]
\centering
\includegraphics[width=0.47\textwidth]{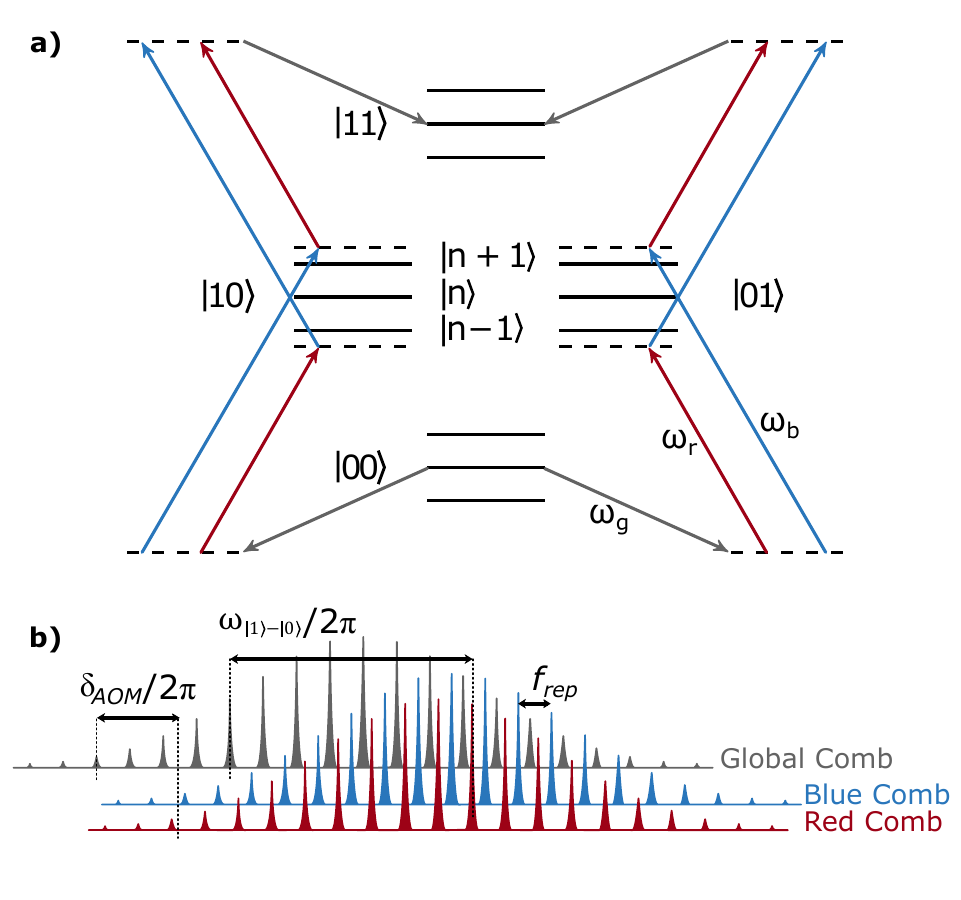}
\caption{Raman transitions and frequency combs to generate $MS(\theta)$. In a), the level structure for the $MS(\theta)$ Raman transitions. $\ket{n}$ is the initial phonon number in a particular motional mode, and red- and blue-sideband transitions are $\ket{n-1}$ and $\ket{n+1}$ respectively. The Raman transitions operate through virtual states to address the qubit motional modes. The grey arrow is the global beam $\omega_g$ acting as one leg of the transition, while the individual beams complete the transition as symmetric red-(blue-) detuned transitions $\omega_r$ ($\omega_b$). As the qubits never occupy the virtual intermediate states, the population oscillates between $\ket{00}$ and $\ket{11}$. b) A graphical representation of the three frequency combs involved in the $MS(\theta)$ gate. $f_{rep}$ indicates the repetition rate of the laser, $\delta_{AOM}/2\pi$ represents the median frequency shift of the red and blue combs relative to the global comb. These are then shifted further from $\delta_{AOM}/2\pi$ by $\pm\delta_c/2\pi$ (not denoted here). To generate the necessary transitions to drive the gate near $\omega_{\ket{1}-\ket{0}}/2\pi$, tooth $j$ of the global comb and teeth $j+105$ of the red and blue combs combine. Comb tooth separation of 105 is represented by a smaller number of comb teeth in the figure for graphical purposes.}
\label{fig:MSandComb}
\end{figure}

However, in our implementation of the MS gate, we split a mode locked laser (repetition rate $f_\text{rep} = 120.125\,$MHz, and pulse duration $\tau_{pulse} = 3.9\,$ps) into two paths, the global and IA paths. Using AOMs, different frequency shifts are produced on each of the paths. 
To drive the MS gate, we need two separate Raman transitions as shown in Fig~\ref{fig:MSandComb}a. The frequency on global beam $\omega_g$ is shared between these transitions. To complete the gate, red- and blue- detuned sidebands are needed, and these tones $\omega_b$ and $\omega_r$ are both applied to the IA AOM.

By definition, a pulsed laser is also a frequency comb, thus each of those tones has a corresponding comb offset from each other, shown in  Fig~\ref{fig:MSandComb}b~\cite{clark2021}. 
The angular frequencies of the $j$-th comb tooth of each comb are offset from one another,
\begin{align}
    \omega_b^{(j)} &= \omega_g^{(j)} + \delta_{AOM}+\delta_c,\nonumber\\
    \omega_r^{(j)} &= \omega_g^{(j)} + \delta_{AOM}-\delta_c,
\end{align}
in which $\delta_{AOM}/2\pi \approx30$ MHz is the difference between the singular frequency applied to the global beam AOM and the median frequency applied to the IA beam AOM. We can define $\omega_g^{(j)}$ in terms of the frequency comb:
\begin{align}
    \omega_g^{(j)} &= \omega_{SP} + \Delta + j2\pi f_\text{rep}, 
\end{align}
where $\omega_{SP}/2\pi$ is the $^{2}S_{1/2}\,\ket{F=1, m_F=0} - ^{2}P_{1/2}\, \\ \ket{F=0, m_F=0}$ transition frequency, and $\Delta$ is the single-photon detuning of tooth zero of comb $g$ from this transition.  To drive the MS gate, our Raman transition needs to have a frequency difference of approximately $\omega_{\ket{1}-\ket{0}}/2\pi \approx 12.643$\,GHz. This is realized with comb teeth separated by 105 harmonics ($105\times 120.125\,$MHz $ \approx 12.613\,$GHz) plus the additional shift $\delta_{AOM} \pm \delta_c$. 

To be precise, the the $j ^{th}$ tooth of comb $g$ combines with the $j+105^{th}$ teeth of combs $r$ and $b$  to primarily drive the blue and red motional sidebands of the two-photon Raman transition between qubit states, respectively. As described below, these frequency combs create a significant contribution to the fourth-order light shift~\cite{lee2016lightshifts}, which we need to account for in our calibration routine.

We describe the resonant $^{2}S_{1/2} - ^{2}P_{1/2}$ Rabi rate of the $j^{th}$ tooth of each frequency comb $\alpha = g, b, r$ by a hyperbolic secant envelope~\cite{Mizrahi2013},
\begin{equation}
    h_\alpha^{(j)} = h_\alpha^{(0)} \sech(j 2\pi f_\text{rep} \tau_\text{pulse}),
\end{equation}
where the single-photon Rabi rate $h_\alpha^{(0)}$ corresponds to tooth $j=0$ of comb $\alpha$. 

The resonant two-photon Rabi rate of the qubit transition generated from tooth $j$ of comb $\alpha$ and tooth $j+l$ of comb $\beta$ is,
\begin{equation}
\begin{split}
    \Omega_{\alpha, \beta}^{(l)}&= \sum_j \frac{h_\alpha^{(j)}h_\beta^{(j+l)}}{2}\\
    &\times\left(\frac{1}{ \Delta + j2\pi f_\text{rep}} - \frac{2}{\Delta + j2\pi f_\text{rep} - \omega_{PP}} \right),
    \label{eq:rabi_rate_b}
\end{split}
\end{equation}
where $\omega_{PP}$ is the fine-structure splitting of the $^{2}P_{3/2}$ and $^{2}P_{1/2}$ levels. In this equation, we have neglected the hyperfine splittings of the $P$ states and the inter-comb detunings $\delta_{AOM} \pm \delta_{c}$ which are all much smaller than $\Delta$ and $\Delta - \omega_{PP}$. We have also assumed that the laser polarization of all beams is $\hat{\sigma}_{\pm}$ which maximizes the Rabi rate of the qubit transition and prevents $\ket{F=0, m_F=0} - \ket{F=1, m_F=\pm 1}$ transitions~\cite{lee2016lightshifts}. In practice, the laser polarization of the global beam is optimized to maximize the Rabi rate of a counter-propagating beam, but as the IA beams must be able to generate both co-propagating and counter-propagating gates, their polarization will not be purely $\hat{\sigma}_{\pm}$.

The fourth-order light shift from combs $\alpha$ and $\beta$ on each qubit level is,
\begin{equation}
\label{eq:ls}
\Delta E_{\alpha,\beta}^{(\pm)} = \pm \sum_l \frac{\left(\Omega_{\alpha, \beta}^{(l)}\right)^2}{4(\omega_\alpha^{(l)} - \omega_\beta^{(0)} - \omega_{\ket{1}-\ket{0}})},
\end{equation}
where the $+$ and $-$ sign correspond to a shift of level $\ket{1}$ and $\ket{0}$, respectively. Hence, the differential fourth-order light shift on the qubit transition from each pair of combs is $\delta^\text{LS}_{\alpha,\beta} = \Delta E_{\alpha,\beta}^{(+)} - \Delta E_{\alpha,\beta}^{(-)} = 2\Delta E_{\alpha,\beta}^{(+)}$. The inter-comb detunings are important for determining the value of Eq.~\ref{eq:ls} as they dictate the two-photon detuning from the qubit transition for pairs of comb teeth separated by $l$ harmonics and thus cannot be neglected as in Eq.~\ref{eq:rabi_rate_b}.

An example $MS(\pi/2)$ gate (with target qubits $q_0$ and $q_1$ in a 2- or 4-qubit register) has standard parameters $\tau=250\,\mu$s and a blue-sideband detuning of $\delta_{k} = \omega_b^{(105)} - \omega_g^{(0)} - \omega_{\ket{1}-\ket{0}}-\nu_0 = 52\,$kHz from the lowest motional mode $k=0$. This gate requires a two-photon Rabi rate of $\Omega_{b, g}^{(105)}/2\pi = \Omega_{r, g}^{(105)}/2\pi = 122.1\,$kHz, which we achieve by scaling the value of $h^{(0)}_\alpha$ (for each $\alpha$) accordingly. We approximate beam amplitudes and polarizations of the experiment in the model. Specifically, the global beam has $\sim2\times$ the amplitude of the individual beams at the ion. Likewise, the global beam polarization is set to maximize counter-propagating interactions at the expense of co-propagating interactions for which we measure a $7\times$ reduction in the amount of $\hat{\pi}$ polarization (driving co-propagating) relative to $\hat{\sigma}_{\pm}$ polarization (driving counter-propagating). Therefore we reduce the fourth-order shifts from the global beam intra-comb (a co-propagating interaction) by a factor of $7^2$ due to incompatible polarizations to induce a light shift. This results in a total differential fourth-order light shift approximately 
$\sum_{\alpha,\beta}\delta^\text{LS}_{\alpha,\beta} = 418$Hz. This light shift is clearly dependent on the overall power in each comb, changing for different $\theta$ and $h_{\alpha}^{(j)}$, and thus methods for mitigating the light shift must account for these dependencies.

\subsection{Empirical Cancellation of Light Shifts} 
Fourth-order light shifts from the red or blue Raman drive alone cause the qubit to quickly dephase; therefore, it is important that these terms cancel to improve the coherence of the MS($\theta$) gate. In the previous section, we considered the effect of all combs simultaneously. If instead, we examine the effects of $\delta^\text{LS}_{g,b}$ and $\delta^\text{LS}_{g,r}$ separately, we find they are of opposite sign and comparable magnitudes. Instead of scaling $h_{\alpha}^{(j)}$ for each comb together, we can scale each comb independently such that $\delta^\text{LS}_{g,b}$ and $\delta^\text{LS}_{g,r}$ cancel, but with the overall two-photon Rabi rate being comparable to the target. 

To find the optimal cancellation point, we begin by directly measuring the resulting light shift and its coherence via a Ramsey measurement. Specifically, we insert increasing numbers of `single-qubit' $MS(\pi/2)$ gates between two microwave $\pi/2$ pulses. In this case, we drive an $MS(\pi/2)$ gate with the parameters necessary for the maximally entangling two-qubit version of the gate but only with a single ion's IA beam on. 

In this way, we prevent any changes to the spin probabilities thus isolating the magnitude of the light shift. We repeat this scan for different ratios $a_{\rm blue}/a_{\rm red} = \zeta_{\rm br}$ of the red and blue MS gate tones applied to the AOMs. While the ratios are varied, the amplitudes are changed in order to maintain the parameters for a two-qubit $MS(\pi/2)$ gate. We note that the scaling of $\zeta_{\rm br}$ is performed in a manner consistent with the amplitude scaling necessary for arbitrary $\theta$ -- specifically, only the amplitude of the global beam is adjusted to determine rotation, thus leaving adjustment on the IA tones for the purposes of balancing $\zeta_{\rm br}$ for the light shift.

\begin{figure}[h]
\centering
\includegraphics[width=0.47\textwidth]{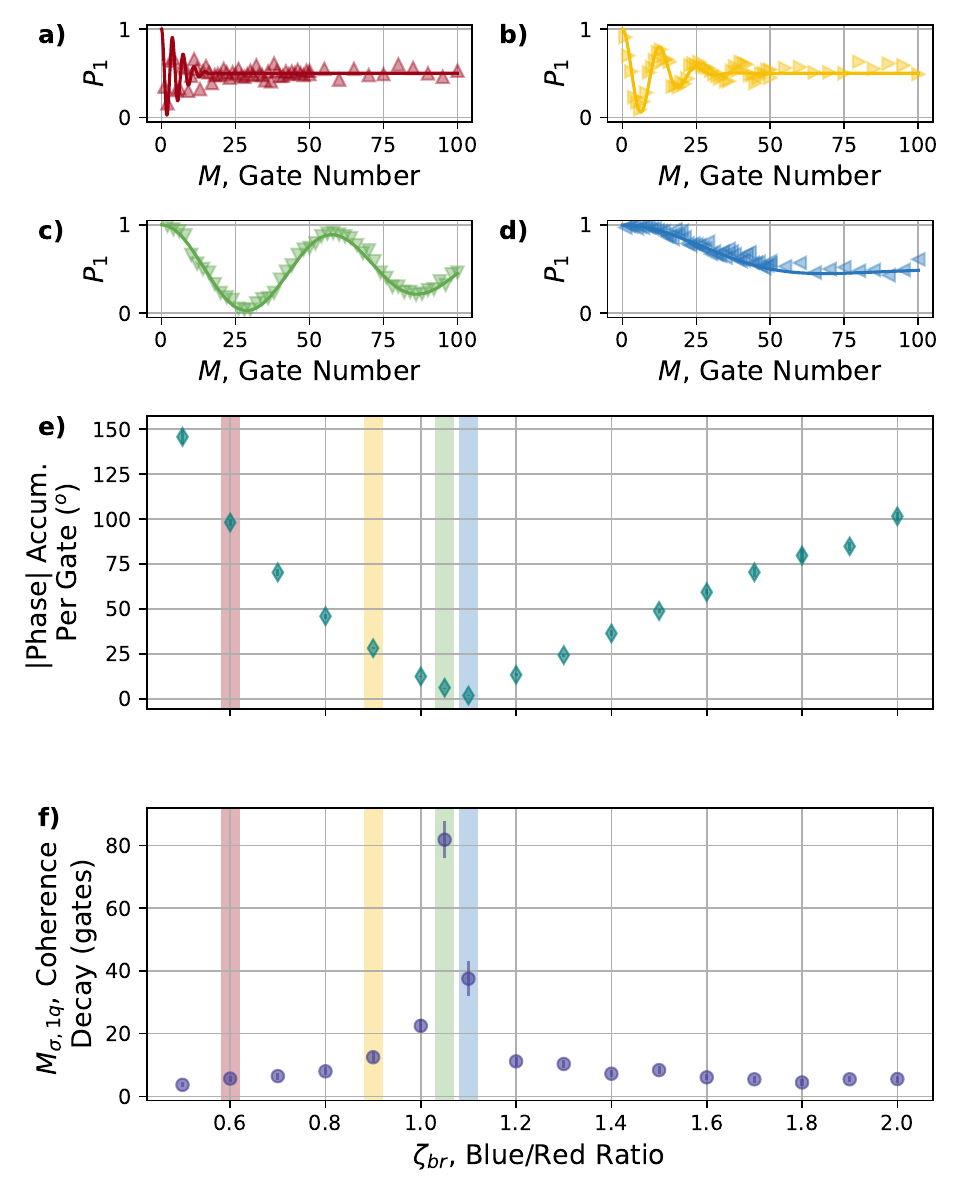}
\caption{Light shift as function of blue/red amplitude ratio ($\zeta_{\rm br}$). a-d) Ramsey measurements with $\zeta_{\rm br} = 0.6, 0.9, 1.05, 1.1$, respectively. f) The accumulated phase per gate (teal diamonds) g) and coherence decay constants (blue circles) were extracted from Ramsey measurements for various blue/red ratios, using a cosine with a Gaussian decay profile. Each of the shaded regions correspond to data extracted from plots a-d.}
\label{fig:microwaveramsey}
\end{figure}

Figures ~\ref{fig:microwaveramsey}a-d show these Ramsey measurements after $M$ applications of a single-qubit $MS(\pi/2)$. We quantify the light shift effects by fitting the Ramsey measurements to a cosine with a Gaussian decay profile to find the accumulated phase and coherence decay constants.

Figure~\ref{fig:microwaveramsey}a shows the Ramsey reasurement for a significant mismatch of $\zeta_{\rm br}=0.6$. The accumulated phase is $\sim98\degree$ per single-qubit $MS(\pi/2)$ gate, along with a fast Gaussian decay constant of $M_{\sigma, 1q} = 5.7$ gates which is highlighted in red on Fig.~\ref{fig:microwaveramsey}e and f. Figures~\ref{fig:microwaveramsey}e,f also show that as $\zeta_{\rm br}$ is increased, the degree of cancellation improves, evidenced by the decreasing phase accumulation per gate and the increasing coherence of Ramsey oscillations. Interestingly, there is a slight difference between the ratio at which accumulated phase is minimized and the ratio at which the coherence is maximized. The longest coherence occurs at $\zeta_{\rm br} = 1.05$ and is found to have $M_{\sigma, 1q} = 81.9$ gates with each MS gate accumulating $6.2\degree$ of light shift (Fig.~\ref{fig:microwaveramsey}c and the green shaded regions in Fig.~\ref{fig:microwaveramsey}e and f). However, the phase is best nullified at $\zeta_{\rm br} = 1.1$ with $M_{\sigma, 1q} = 37.5$ and each gate accumulating $1.8\degree$ of light shift (Fig.~\ref{fig:microwaveramsey}d and the blue shaded regions in Fig.~\ref{fig:microwaveramsey}e and f). To summarize, the ratio $\zeta_{\rm br}$ resulting in the most coherent operation leaves a residual light shift, which we compensate with a calibration described in the next section.
\begin{figure}[h]
\captionsetup[subfigure]{labelformat=empty}
    \begin{subfigure}[]{0.47\textwidth}
        \centering
        \begin{resizebox}{1\textwidth}{!}{
            \Qcircuit @R=.35em {
            & \gate{{R}_y^{\mu}(\frac{\pi}{2})} & \gate{{MS}_1(\frac{\pi}{2})} & \qw & \gate{{R}_y^{\mu}(\pi)} & \gate{{MS}_1(\frac{\pi}{2})} & \qw & \gate{{R}_y^{\mu}(\frac{\pi}{2})}  & \qw  \\
            & \gate{{R}_y^{\mu}(\frac{\pi}{2})} & \qw & \gate{{MS}_2(\frac{\pi}{2})} & \gate{{R}_y^{\mu}(\pi)} & \qw & \gate{{MS}_2(\frac{\pi}{2})} & \gate{{R}_y^{\mu}(\frac{\pi}{2})}  & \qw               \gategroup{1}{7}{2}{7}{1.5em}{)} 
             \gategroup{1}{3}{2}{3}{1.5em}{(}
            \\
         }
         }
         \end{resizebox}
        \caption{}
        \label{fig:echo-diagram}
    \end{subfigure}
    \centering
    \includegraphics[width=0.47\textwidth]{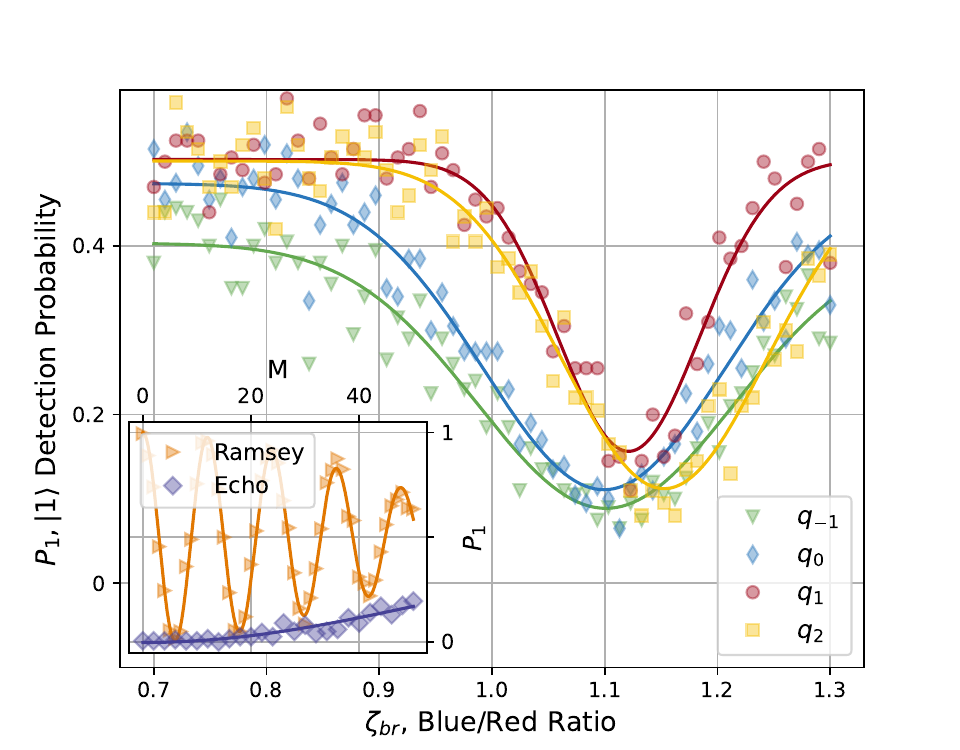}
    \caption{Microwave echo calibration of $\zeta_{\rm br}$ for selecting point of best coherence. A two-qubit pulse sequence example is shown on the top where we start by applying a microwave $R_y^{\mu} (\frac{\pi}{2})$ pulse to each ion. We test the echo sequence against the Ramsey sequence by applying the equivalent single-qubit $MS(\pi/2)$ gates as in the Ramsey sequence on each of the ions in the chain, separated in time such that the light drives only one ion at a time. These are separated by an echoing microwave $\pi$ pulse, reversing the phase evolution of the spins. Then another set of single-qubit MS gates are applied. Finally, a final projection pulse is applied via another microwave $R_y^{\mu}(\frac{\pi}{2})$ gate.  In the bottom, plotted are the results of the echo calibration of $\zeta_{\rm br}$ taken on four qubits, with each qubit's trace fit to a Gaussian. For each qubit [$q_{-1}$,$q_0$,$q_1$,$q_2$], the best $\zeta_{\rm br}$ is found to be [1.10, 1.10, 1.12, 1.15]. The inset shows the a Ramsey sequence and the equivalent echo sequence on one qubit, with both showing similar decay profiles.
    }
    \label{fig:microwaveecho}
\end{figure}

In order to more easily calibrate $\zeta_{\rm br}$ for any given ion and gate pair, we test an echo sequence against the Ramsey sequence and find similar decay profiles (Fig.~\ref{fig:microwaveecho}) indicating largely homogeneous dephasing on these timescales. Thus for calibration, we can now simply scan $\zeta_{\rm br}$ within an echo containing a fixed number of single-qubit MS gates and select the point where the ion best returns to its initial state. For a chain of ions, this calibration is done within a single scan by applying a single-qubit MS gate pulse to each ion in series (only one ion at a time to avoid driving any actual entangling gates). An example set of data for this calibration is shown in Fig.~\ref{fig:microwaveecho} for four qubits. 

\subsection{Residual Light Shift Cancellation}

While calibrating $\zeta_{\rm br}$ eliminates the strongest contributions of the fourth-order light shift, there is still a residual light shift. As shown in Fig.~\ref{fig:microwaveramsey}, this residual light shift is $6\degree$, but we typically find it varies across different MS gate pairs in a larger chain, and can be anywhere from $3-30\degree$ for maximally entangling gates. To correct for this residual light shift we take advantage of a ``frame rotation,'' a phase-bookkeeping approach which advances the phase on all subsequent pulses on any given qubit.
    
We describe the nuances of how this frame rotation is implemented in our control hardware, Octet, in detail in~\cite{clark2021}, but relevant features are discussed here. In particular, across a circuit, these frame rotations are tracked within the control hardware for each qubit. To be clear, we refer to the overall tracked phase per qubit ($i$) per circuit as the ``qubit frame,'' $\Phi_{i}$ and the individual operations that modify that frame as ``frame rotations,'' $\phi_{f, i}(t)$. We note that $\phi_{f, i}(t)$ can be programmed to occur during a gate or as a standalone operation, i.e. a virtual Z (or phase) gate. In the context of the Raman transitions that underly our gates, phases are defined as the relation between certain legs of the specific Raman transitions, and thus care must be taken to ensure the qubit frame is referenced properly. For instance, in the MS gate, we reference the qubit frame $\Phi_{i}$ on both tones of the IA beam.
    
\begin{figure}[h]
\centering
\includegraphics[width=0.47\textwidth]{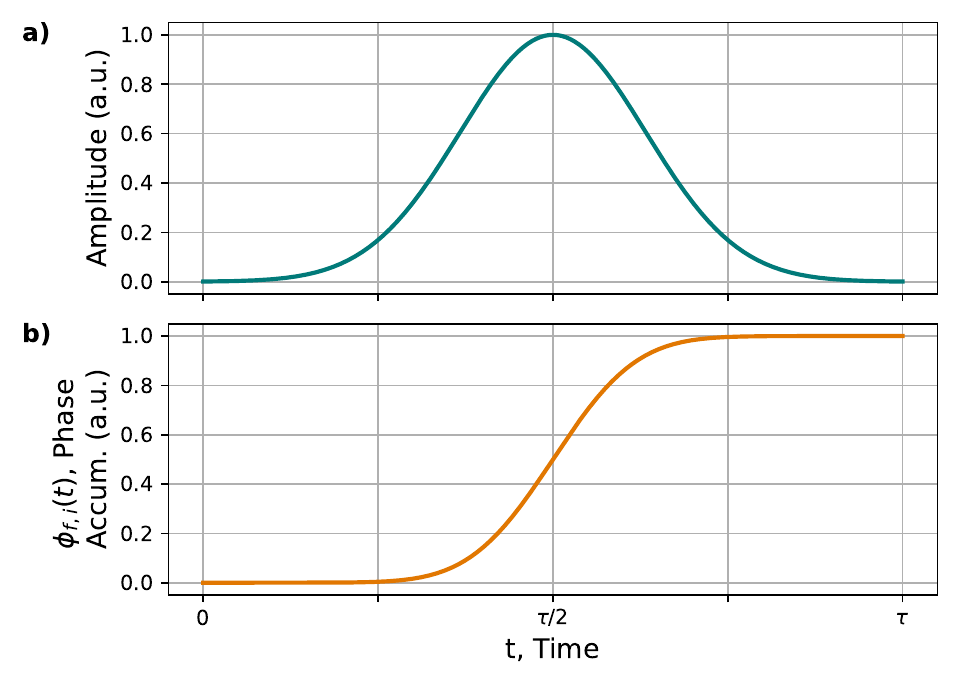}
\caption{ Pulse shaping for amplitude and frame rotation of the $MS(\theta)$ gate, a) The amplitude of the waveform (teal) follows a Gaussian with $\sigma = 0.133\tau$, while b) the frame rotation $\phi_{f,i}(t)$ follows the integral of the square of Gaussian amplitude, $\erf(\sqrt{2}t)$ (orange), as the residual light shift is predominantly fourth-order.
}
\label{fig:gauss_and_erf}
\end{figure}

Harnessing this the control hardware capability, we describe an implementation that cancels the residual light shift. We use $\phi_{f, i}(t)$ as a \textit{dynamic} phase shift concurrent with the MS gate in order to counteract the influence of the light shift during the varying amplitude pulse (Gaussian-shaped). Since the residual light shift is predominantly due to fourth-order shifts, it is proportional to the square of the Rabi rate. This means the light shift will cause the gate to accumulate phase as the integral of the square of that Gaussian envelope, which is a scaled error function, $\erf(\sqrt{2}t)$ (see Fig.~\ref{fig:gauss_and_erf}b). We therefore program the frame rotation $\phi_{f, i}(t)$ to also accumulate phase as $\erf(\sqrt{2}t)$ spanning the duration of the gate, but with the opposing sign of the phase from the light shift in order to cancel it.

\begin{figure}[h!!!!!]
\centering
\includegraphics[width=0.47\textwidth]{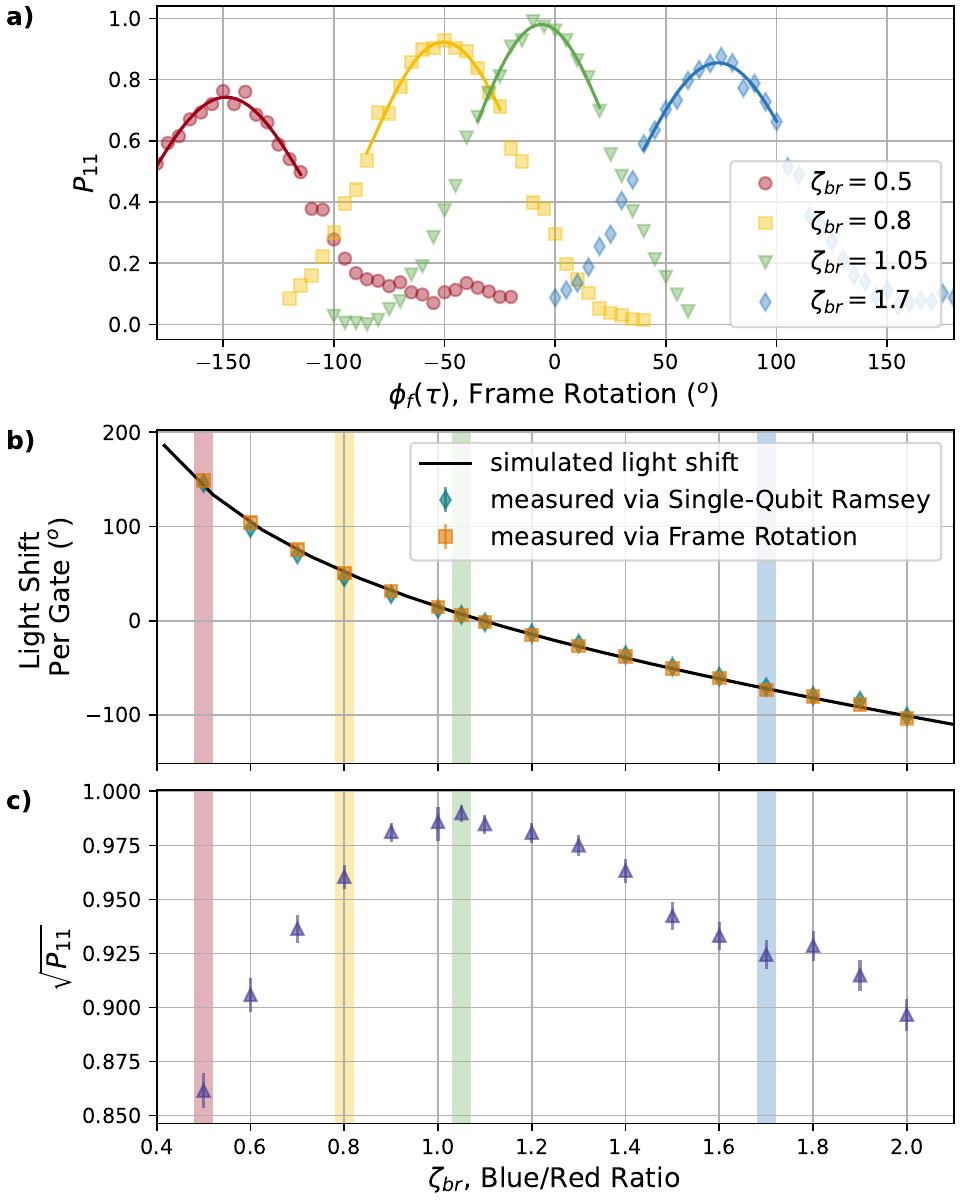}
\caption{Different impacts of the light shift as a function of $\zeta_{\rm br}$ measured with the 2x $MS(\pi/2)$ gates. a) The probability in $\ket{11}$ after two sequential $MS(\pi/2)$ gates vs. $\phi_{f,i}(\tau)$. The four different data sets have different values of $\zeta_{\rm br}$.  b) From a), the center of each set is fit to extract the frame rotation needed to cancel the residual light shift (orange squares), i.e. the sign of the measured frame rotation is inverted to indicated the light shift that is being canceled. The single-qubit $MS(\pi/2)$ calibration results in Fig.~\ref{fig:microwaveramsey}e have been reproduced (teal diamonds) and assigned the appropriate directionality (sign) as the original measurement is unable to determine the sign.  The frame rotation and the single-qubit $MS(\pi/2)$ Ramsey measurements were interleaved to reduce the effect of system drift. A simulation (black curve) based of the fourth-order light shift resulting from the effect of the other pulsed laser comb harmonics is fit to the data to account for imperfect polarization and mismatch between the intended and actual $\zeta_{\rm br}$ at the ion. c) An estimate of the gate performance, $\sqrt{P_{11}}$ after two applications of $MS(\pi/2)$, as determined from the frame rotation measurements reveals significant degradation as $\zeta_{\rm br}$ moves away from the ideal 1.05. Error bars in b) and c) are $2\sigma$ confidence intervals either determined from a maximum likelihood estimation or basic curve fitting routine.}
\label{fig:framerotvsramsey}
\end{figure}

To calibrate the magnitude of the correction needed, we perform two sequential $MS(\pi/2)$ gates on an initial state of $\ket{00}$, which should result in complete population transfer to $\ket{11}$. However, in the presence of light shifts, the second gate will not be aligned in phase with the final state of the first, resulting in less overall population transfer to $\ket{11}$. Here, we vary the total accumulation of phase of the frame rotation, which is $\phi_{f, i}(t)$ at $t = \tau$, or more compactly $\phi_{f, i}(\tau)$. We then find $\phi_{f, i}(\tau)$ which corresponds to the maximal $\ket{11}$ population, as shown in the green data of Fig.~\ref{fig:framerotvsramsey}a. During calibrations, we fit the entirety of the scan with a simple Gaussian curve fitting routine, taking its center to be the calibrated frame rotation.  However, the true functional form is not a Gaussian, and so for more thorough investigations and fidelity estimations, we utilize a maximum likelihood estimation that fits a Gaussian only to the upper half of the curve to extract the amplitude (for fidelity estimations) and the center (for frame rotation calibrations), demonstrated in Fig.~\ref{fig:framerotvsramsey}a. When we compare the two different methods to determine the necessary frame rotation, we find they are typically within $2\degree$ of one another (a example set of 22 calibration measurements yielded an average difference of $0.95\degree$ or $\sim1.5\%$ of the Gaussian 2$\sigma$). 

For calibrations, we determine $\zeta_{\rm br}$ prior to calibrating the frame rotation. To confirm that the residual shift canceled through this method matches our previous measurements of phase accumulation (Fig.~\ref{fig:microwaveramsey}), we investigate how $\zeta_{\rm br}$ affects the determination of the best $\phi_{f,i}(\tau)$ in Fig.~\ref{fig:framerotvsramsey}. Four examples of the $\phi_{f,i}(\tau)$ calibration are presented, each at taken at a different $\zeta_{\rm br}$. Each $\zeta_{\rm br}$ yields a different necessary $\phi_{f,i}(\tau)$ for cancellation of the light shift. When compared to the light shift imparted by the single-qubit $MS(\pi/2)$ as a function of $\zeta_{\rm br}$ (originally presented in Fig.~\ref{fig:microwaveramsey}e and reproduced as teal diamonds in Fig.~\ref{fig:framerotvsramsey}b), there is significant agreement. This agreement further indicates the measured necessary frame rotation does indeed cancel any light shift of the gate. Frame rotation scans presented in Fig.~\ref{fig:framerotvsramsey} were performed interleaved with the single-qubit Ramsey measurements in Fig.~\ref{fig:microwaveramsey} to combat possible drift in the system.

While the frame rotation alone is sufficient to cancel the entire light shift, there are clear performance impacts for non-ideal $\zeta_{\rm br}$ as shown in Fig.~\ref{fig:framerotvsramsey}a. At $\zeta_{\rm br} = 1.05$, nearly all the population is transferred to $\ket{11}$ at $\phi_{f,i}(\tau) = -6.25\degree$ (which cancels a light shift of $+6.25\degree$). However, as $\zeta_{\rm br}$ moves away from its optimal point so does the maximal degree of population transfer to $\ket{11}$. As such, we can also extract a rough performance estimate of the gate based on its ability to generate $\ket{11}$. To estimate the performance of a single gate, we take the square root of the probability of measuring $\ket{11}$, or $\sqrt{P_{11}}$. When we plot that performance metric as a function of $\zeta_{\rm br}$ in Fig.~\ref{fig:framerotvsramsey}c, there are clear reductions away from the optimal point, $\zeta_{\rm br} = 1.05$, similar to the reduction in coherence seen in Fig.~\ref{fig:microwaveramsey}f.

A simulation of the fourth-order light shift is also plotted in Fig.~\ref{fig:framerotvsramsey}b with the measurements. The light shifts from each combination of comb teeth (i.e. red with global, blue with global, and red with blue as well as intra-comb combinations) are summed (and scaled based on empirically estimates that approximate beam amplitudes and polarizations) to determine the simulated fourth-order contribution. We fit the simulation to the empirical results, and find the Rabi rate to be 1.10 times the measured rate (at $\zeta_{\rm br} = 1.0$, we measure $\Omega_{b, g}^{(105)}/2\pi = \Omega_{r, g}^{(105)}/2\pi = 122.1$ kHz), and $\zeta_{\rm br}$ to be 1.04 times the intended (or programmed) ratio. These mismatches between the measurements and simulations are not unexpected. The mismatch in Rabi rate is likely due to imperfect polarizations at the ion not adequately captured in the model. Specifically, $\hat{\pi}$ polarization components of the light are unable to drive counter-propagating carrier Raman transitions, but will contribute to light shifts from the $\ket{F=0, m_F=0} - \ket{F=1, m_F=\pm 1}$ Zeeman transitions (separated by $\sim5.96$MHz from the carrier) which are neglected in this model. Likewise, the minor mismatch in the amplitudes we program relative to the amplitudes at the ion are a result of slight differences in AOM efficiencies at the blue- and red-detuned frequencies.
    
\begin{figure}[t]
\centering
\includegraphics[width=0.47\textwidth]{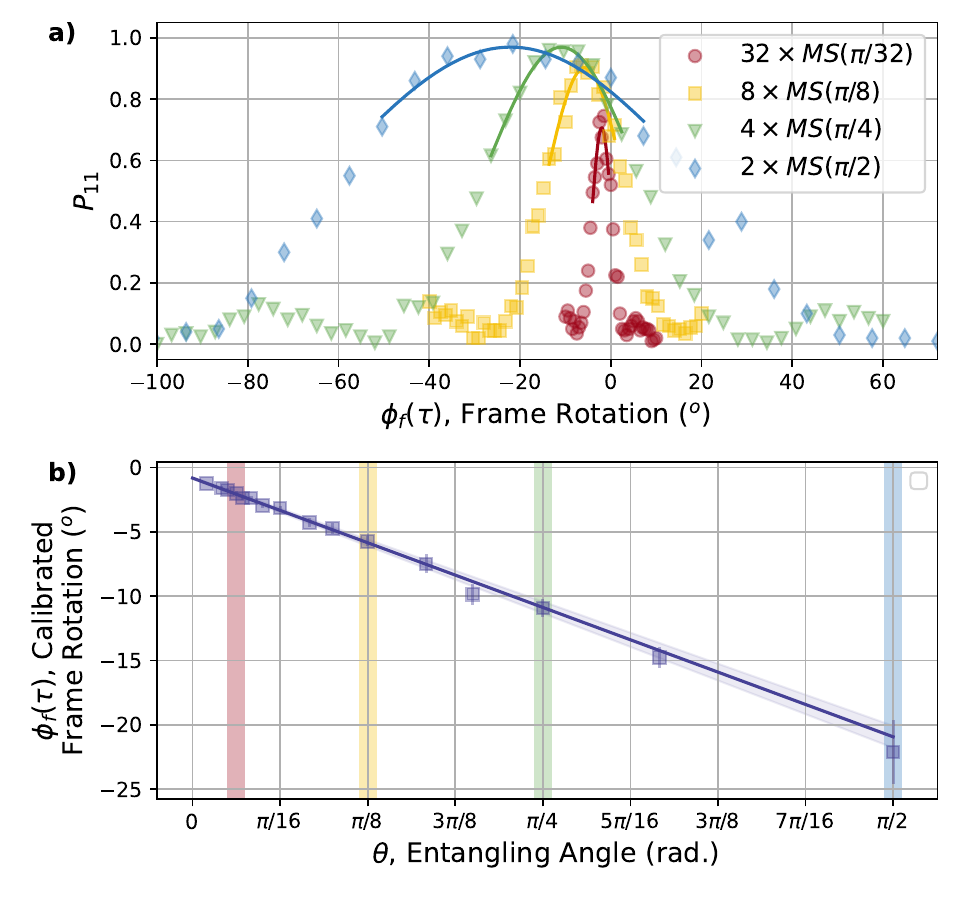}
\caption{Frame rotations for various different $MS(\theta)$  a) $M \times MS(\pi/M)$ are performed while the total frame rotation magnitude, $\phi_{f,i}(\tau)$ is scanned. Plotted are examples (M = [2,4,8,32]) from the complete dataset. The peak of the upper half of the distribution is fit for each M in the dataset to a Gaussian through maximum likelihood estimation and plotted in b) where the shaded regions correspond to the examples presented in a). There is a clear linear relationship between $\theta$ and the $\phi_{f,i}(\tau)$. This relationship is used to interpolate the needed $\phi_{f,i}(\tau)$ for any given $\theta$.} 
\label{fig:framerot}
\end{figure}
    
The next challenge is understanding how to scale the magnitude of these light shifts for any arbitrary entangling angle $\theta$. In this case, we use $M$ successive $MS(\pi/M)$ pulses and repeat the procedure above for a variety of $M$. In Fig.~\ref{fig:framerot}a, we see that with decreasing $\theta$, the residual light shift to be canceled decreases as well. Interestingly, we find a linear relationship between $\theta$ and the frame rotation, $\phi_{f,i}(\tau)$ needed to cancel the light shift, as shown in Fig.~\ref{fig:framerot}b. Therefore, in practice, we only need to find two points along this functional form to interpolate/extrapolate the needed frame rotation for any arbitrary rotation. We find $2\times MS(\pi/2)$ and $32\times MS(\pi/32)$ are sufficient.

\section{ZZ Gates and Phase Agnosticism} \label{sec:zz}

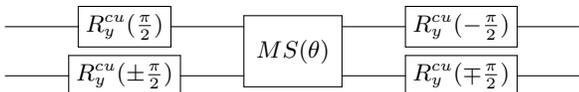
\begin{figure}[h]

    \Qcircuit @R=.35em {
        & \gate{{R}_y^{cu}(\frac{\pi}{2})} & \multigate{1}{{MS}(\theta)} &  \gate{{R}_y^{cu}(-\frac{\pi}{2})}  & \qw  \\
        &  \gate{{R}_y^{cu}(\pm\frac{\pi}{2})} & \ghost{{MS}(\theta)} &  \gate{{R}_y^{cu}(\mp\frac{\pi}{2})}  & \qw\\
        \vspace{3cm} \\
    }
    \caption{$\zz(\theta)$ circuit. To convert the $MS(\theta)$ to the $ZZ(\theta)$, the $MS(\theta)$ is surrounded by counter- propagating wrapper gates, $R_y^{cu}(\pm\pi/2)$. If the desired $\theta$ is positive (negative), the sign of the wrapper gates on both qubits are matched (mismatched). The waveforms for the internal $MS(\theta)$ are performed at a specific phase relation to always generate an $XX(|\theta|)$ gate (i.e. 0 ($\pi$) for a native $XX$ ($-XX$) interaction)} \label{fig:zzcircuit} 
\end{figure}

The final step in our procedure is to convert the $MS(\theta)$ gate into its phase-agnostic $ZZ(\theta)$ formulation~\cite{lee2005zz, baldwin2020zz, chow2024} for integration into larger and deeper circuits. We note that $ZZ(\theta)$ gates are phase agnostic because single-qubit Z gates commute with $ZZ(\theta)$. The rationale for this conversion is two-fold. To begin, two-qubit $MS(\theta)$ gates are performed in a counter-propagating configuration (two tones on the IA beam and one tone on the global beam). Our single-qubit gates can be performed either in a counter-propagating configuration or a co-propagating configuration (two tones on the IA beam), but in many cases are performed in a co-propagating configuration as they are not motionally sensitive. However, phase instabilities arise due to path length differences when naively combining co- and counter-propagating gates. Thus, we take advantage of \textit{counter-propagating} single-qubit gates to perform a change of basis whenever a two qubit-gate is needed. Specifically, this allows us to transform the XX-type interaction to a ZZ interaction using counter-propagating ``wrapper'' gates, $R_y^{cu}(\pm\pi/2)$ to surround the $MS(\theta)$ interaction, as shown in Fig.~\ref{fig:zzcircuit}~\cite{lee2005zz}.

We also find that this approach eliminates errors in $\theta$ that arise from phase-dependent optical crosstalk on nearest and next-nearest neighbor pairs~\cite{chow2024}. These errors are dependent on the phase relationship of the \textit{waveforms} generating an $MS(\theta)$ gate. Because most circuits on the QSCOUT system rely on the ubiquitous use of frame rotations (both as the dynamic phase shifts for light shift cancellation described in the previous section, and as programmed virtual Z gates) this means that all qubit frames, $\Phi_{i}(t)$, within a circuit at any given time may vary significantly from qubit to qubit. Therefore, for an arbitrary $MS(\theta)$ gate within a given circuit, the actual phase relation between the \textit{waveforms} of the gate pulses for each qubit is undetermined until the application of that particular circuit. To be clear, we are making a distinction between the well-defined programmed phase relationships of each gate at the circuit level and what is transpiring on the waveform-generating control hardware which tracks the use of virtual Z gates and frame rotations to update all subsequent waveforms. For instance, given a simple two-qubit circuit in which a virtual $Z(\pi/2)$ is performed on one of the two qubits prior to an XX-oriented $MS(\theta)$ gate, the phase relation of the waveforms being generated during the $MS(\theta)$ gate would actually be the equivalent of a YX-oriented interaction.

Now, by inserting the $MS(\theta)$ gate within wrapper single-qubit gates to transform it to a $ZZ(\theta)$ gate, this new operation is now agnostic to any accumulated phases on the two qubits' respective frames as any phase operations commute through $ZZ(\theta)$. As such, we are free to specify the inner $MS(\theta)$ operation with any phase relation between the respective qubits' waveforms as long as the accompanying wrapper gates are phased appropriately to match. On the QSCOUT system, we always perform the inner $MS(\theta)$ interaction of the $ZZ(\theta)$ gate at a specific phase relation between the waveforms (i.e. 0 ($\pi$) for a native $XX$ ($-XX$) interaction). For more details on how this approach mitigated the errors in $\theta$, see Ref.~\cite{chow2024}, but here we will describe the nuances of how to specify the phase relationship of the waveforms generating that inner $MS(\theta)$ with our Octet control hardware.

Within Octet, for a given qubit $i$, we have access to \textit{two} separate qubit frames that can be tracked, $\Phi^0_{i}$ (or \framezero/) and $\Phi^1_{i}$ (or \frameone/)\cite{LobserJaqalpaw}. In this case, we utilize $\Phi^0_{i}$ to be the default phase bookkeeper for each qubits' frame. The other frame, $\Phi^1_{i}$, becomes a temporary qubit frame that is reset for each implementation of the $ZZ(\theta)$ gate. During the $ZZ(\theta)$ gate interaction, we reset $\Phi^1_{i} = 0$ for both qubits involved in the gate, and both qubits' reference point becomes $\Phi^1_{i}$. Within this temporary frame, the first set of wrapper gates is performed. Then, the $MS(\theta)$ interaction is performed with the appropriate \textit{dynamic} frame rotation, $\phi_{f,i}(t)$ which now accumulates on $\Phi^1_{i}$ for each qubit, such that by the end of the gate, the frame $\Phi^1_{i} = \phi_{f,i}(\tau)$. Finally, the second set of wrapper gates is performed within frame $\Phi^1_{i}$. After completion of the $ZZ(\theta)$ gate, those qubits return to $\Phi^0_{i}$ for frame tracking. The key benefit of this additional frame is that we do not need manual bookkeeping of phases throughout the circuit in order to reset the phase relation for the $ZZ(\theta)$ gate. Instead by having two separate frames, the hardware natively accumulates the appropriate phase for the qubit frame from virtual Z gates on $\Phi^0_{i}$, while $\Phi^1_{i}$ provides a fixed starting phase relationship for waveforms generating $ZZ(\theta)$ gates.

\section{Gate Performance Metrics and Use Cases} \label{sec:fidelity}
With these considerations, we now investigate the performance of $MS(\theta)$. For these purposes, we will assess the performance of $MS(\theta)$ rather than the circuit-level gate, $ZZ(\theta)$, as the performance of the `wrapper' gates will be the same regardless of $\theta$.

\subsection{State Fidelity Estimates}

To determine the fidelity of a single application of the $MS(\theta)$ operation on an initial state $\ket{00}$, we perform two measurements: first a probability measurement of the qubit states after application of the gate and a parity measurements. The parity measurement consists of a gate operation followed by a varied projection $\pi/2$ pulse on each qubit which results in oscillations between even and odd parity state probabilities. A common fidelity estimate metric~\cite{Figgatt2019}, is thus:

\begin{equation}
\begin{split}
F(\theta) = P_{\ket{00}}\cos^2\left(\frac{\theta}{2}\right)
&+P_{\ket{11}}\sin^2\left(\frac{\theta}{2}\right)\\
&+A_{\Pi}\cos\left(\frac{\theta}{2}\right)\sin\left(\frac{\theta}{2}\right)
\label{eq:fidelity}
\end{split}
\end{equation}

Here $P_{\ket{ij}}$ refers to the probability of measuring state $\ket{ij}$, while $A_{\Pi}$ is the fitted contrast of the parity oscillations, discussed in more detail in Appendix~A. In Fig.~\ref{fig:fidelity}, a clear trend is seen that shows reduced infidelity for decreasing entangling angle. This trend is similar to that seen by other groups using arbitrary-angle $MS(\theta)$ gates as well~\cite{moses2023, nam2020}. Due to significant error bars in our estimates, we caution against deducing a specific scaling behavior.

    \begin{figure}[h]
    \centering
    \includegraphics[width=0.47\textwidth]{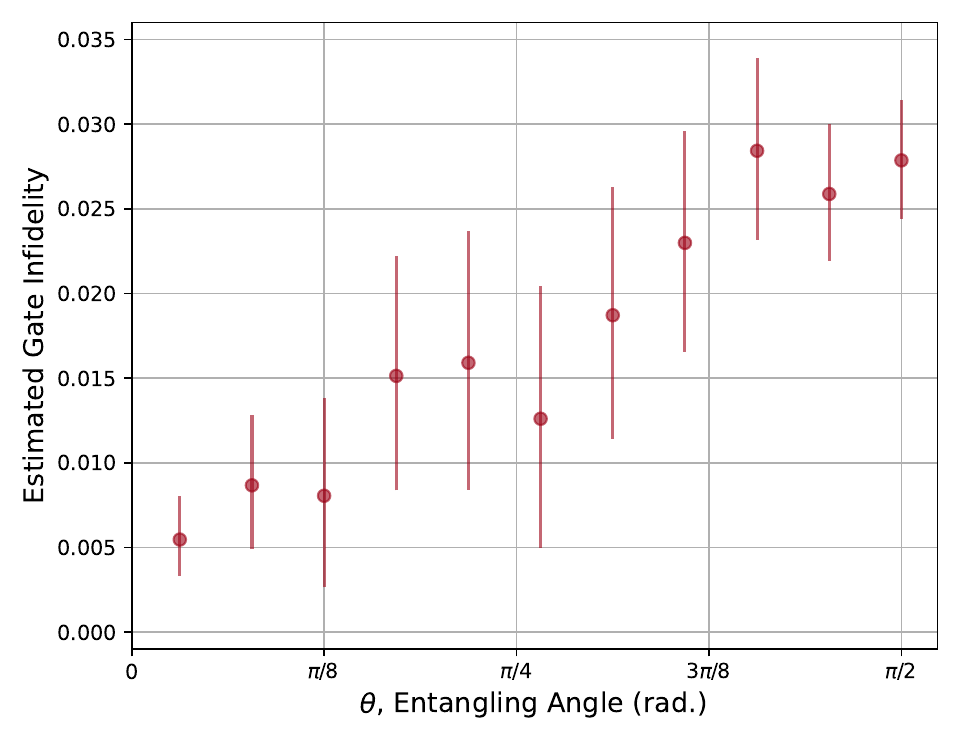}
    \caption{Estimated $MS(\theta)$ infidelity as a function of $\theta$. The infidelity estimate is determined by Eq.~\ref{eq:fidelity}. Error bars are calculated for each independent measurement (the state probabilities and the parity oscillations) and then summed in quadrature. For the probability measurements, $2\sigma$ Wilson scores are used. For the parity measurement, a maximum likelihood estimation is used to fit an amplitude to the parity oscillations, and then the $2\sigma$ confidence intervals are determined from a likelihood estimation performed across a selection of oscillation amplitudes around the fitted amplitude.}
    \label{fig:fidelity}
    \end{figure}

\subsection{What is the best gate angle to use?}

As shown in Fig.~\ref{fig:fidelity}, it is clear that the fidelity \textit{per gate} improves with decreasing $\theta$, and in Fig.~\ref{fig:allmsloops}, the decay constant $M_{\sigma, even}$ increases with decreasing $\theta$. However, the fidelity \textit{per radian} of entangling angle is better for large $\theta$ as seen by the total number of visible Rabi oscillations in Fig.~\ref{fig:allmsloops} improving with increasing $\theta$. By lowering the intensity of the beam for a fixed gate duration to generate smaller $\theta$, we limit the effect of power-dependent errors; however, random frequency shifts and noise in the system will still cause imprecision in both $\theta$ and gate coherence that would compound for multiple applications of small-angle gates when trying to generate a larger-angle gate. 

Therefore, when compiling quantum gates with continuously-parameterized circuits, the minimal sufficient entangling angle to achieve some desired unitary should be used while putting as much of the entangling angle as possible into a single pulse. For example, if a circuit calls for a fully-entangling unitary, a single MS($\pi/2$) should be used, not two applications of MS($\pi/4$). On the other hand, circuits calling for incremental entangling rotations (e.g., small Trotter steps) should use only as large a $\theta$ as necessary to take each step. These conditions are naturally accomplished when using the KAK decomposition for arbitrary SU(4) unitary operations~\cite{tucci2005,campbell2023}, a standard approach used for circuit compilation. Moreover, additional compilation techniques are able to utilize an $MS(\theta)$ gateset for improved performance such as swap mirroring, reordering qubit labels to reduce entangling angle on poorer performing pairs, and circuit approximation~\cite{yale2024superstaq}.   

\section{Conclusion}
\label{sec:conclusion}
Tailoring the degree of entanglement in two-qubit operations offers greater flexibility in circuit design than with a more restricted gateset featuring only maximally entangling gates. This paired with improved gate performance for decreased $\theta$ enables decomposition of arbitary unitary operations into a minimal set of high performance entangling operations. We have detailed the key elements of implementation that enable the realization of arbitrary-angle entangling gates in the QSCOUT system. 

We describe a detailed set of calibrations that result in high performance arbitrary-angle gates. Careful characterization of the saturation and distortion effects on the RF amplitude arising from the hardware is used to precisely tune the amplitude for a desired arbitrary $\theta$. Effects of the fourth-order light shift, namely, phase accumulation during an entangling gate that lead to decoherence, are mitigated by careful selection of the tone amplitudes generating $MS(\theta)$.  We find, however, a residual light shift is still present when selecting tone amplitudes that result in the best entangling gate coherence. To address this, a dynamic virtual phase shift concurrent with the entangling gate is used to nullify the residual light shift phase. 

We have also incorporated a series of approaches that work to provide holistic improvements in $MS(\theta)$ performance. Spectrally compact pulse shaping and motional mode balancing minimize the impact of frequency drift and reduce displacement errors. Additionally, a basis transformation of the entangling interaction to $ZZ(\theta)$ is performed in order to mitigates crosstalk-induced rotation errors and to isolate the gate from phase instabilities arising from different laser propagation paths. 

More broadly, reducing entangling angle shows improved gate performance, both in terms of fidelity and coherence, not only for $MS(\theta)$ gateset on the QSCOUT testbed described here, but also on commercial trapped-ion systems~\cite{nam2020,moses2023}. Likewise, expanding to the circuit compilation, these performance improvements persist at the circuit level~\cite{yale2024superstaq}. Superconducting systems have also begun to take advantage of expanded entangling gatesets to include fractional gates that yield improvements in both performance and circuit depth~\cite{Perez2023, IBMBlog2024}. As such, generating arbitrary amounts of entanglement is an important resource for NISQ processors. 

\iftoggle{ieeeformat}{}{
\section*{Acknowledgements}

This research was supported by the U.S. Department of Energy, Office of Science, Office of Advanced Scientific Computing Research Quantum Testbed Program and by Sandia National Laboratories' Laboratory Directed Research and Development Program. Sandia National Laboratories is a multi-mission laboratory managed and operated by National Technology \& Engineering Solutions of Sandia, LLC (NTESS), a wholly owned subsidiary of Honeywell International Inc., for the U.S. Department of Energy’s National Nuclear Security Administration (DOE/NNSA) under contract DE-NA0003525. This written work is authored by an employee of NTESS. The employee, not NTESS, owns the right, title and interest in and to the written work and is responsible for its contents. Any subjective views or opinions that might be expressed in the written work do not necessarily represent the views of the U.S. Government. The publisher acknowledges that the U.S. Government retains a non-exclusive, paid-up, irrevocable, world-wide license to publish or reproduce the published form of this written work or allow others to do so, for U.S. Government purposes. The DOE will provide public access to results of federally sponsored research in accordance with the DOE Public Access Plan. SAND2025-03151O.
}

\appendix

\iftoggle{ieeeformat}{
\appsection{}

\subsection{Calibration Schedule}}{
\appsection{Calibration Schedule}

}
\label{app:calibration}
In this appendix, we outline the typical calibration schedule for both single- and two-qubit gates performed for users of the QSCOUT hardware. While some users will not require the full suite of calibrations and others will require additional calibrations, this is a representative workflow. 

\begin{figure}
    \centering
    \includegraphics[width=0.47\textwidth]{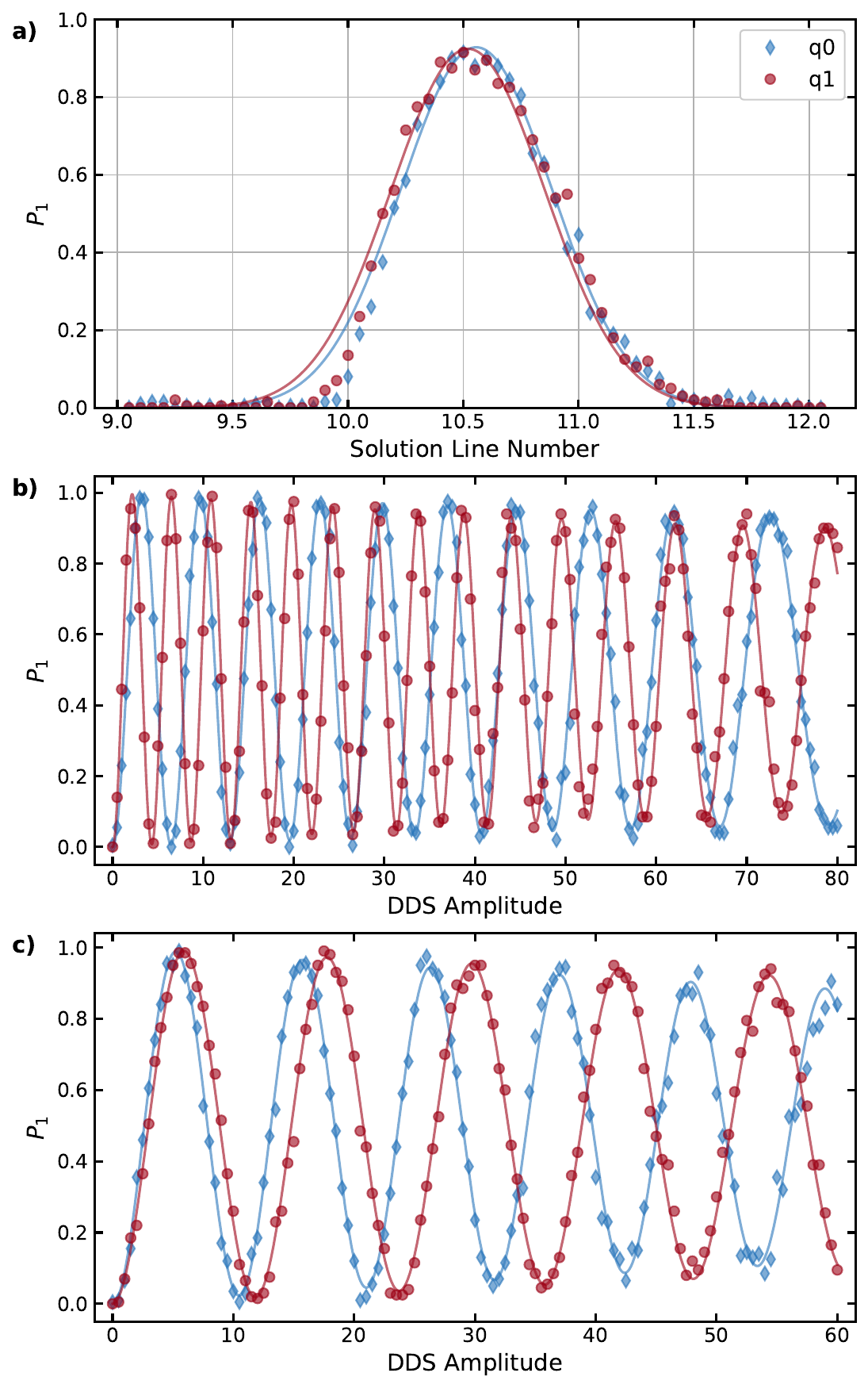}
    \caption{Alignment of individual addressing beams and calibration of individual and counter-propagating RF drive amplitude for qubits $q_0$ (blue) and $q_1$ (red). a) The ion position is swept across the individual addressing beams by tuning the axial trap potential. The x-axis represents line numbers of the solution which are nominally 1 $\mu$m apart. b) and c) Scan of the DDS RF drive amplitude for a single tone on the IA beam for co-propagating (b) and counter-propagating (c) single-qubit gates. Solid lines are fits to the data.}
    \label{fig:linenum_ampscan_cal}
\end{figure}

After loading a chain of ions, the ion florescence is aligned into a multi-core fiber, where each single core is coupled to separate photomultiplier tubes (PMT) for detection. For each calibration step, we Doppler cool and prepare the ions in $\ket{0}$. The axial ion spacing and chain position are aligned to the 335 nm IA beams using DC control voltages applied directly to the trap, which tune the curvature and position of the harmonic trapping potential. Amplitudes necessary to generate slightly less than $\pi$ pulse (when aligned) are applied to the IA beams while tuning the position of trap potential to achieve maximum transfer to $\ket{1}$. Figure \ref{fig:linenum_ampscan_cal}a shows a scan of the axial potential well position which moves 1 $\mu$m for each integer step of the solution line number. Submicron positioning is achieved by interpolating between line numbers. In the scenario that the well position for maximum transfer to $\ket{1}$ does not coincide for all ions, the ion spacing is adjusted by scaling the trap frequency. However, for ion chain lengths of 4 to 6, not all ions will be fully aligned due to the harmonic nature of the potential; therefore, the potential is adjusted such that some ions are intentionally misaligned to have the greatest average coverage for all ions. At ion numbers greater than 6, quartic terms and/or additional spectator ions may be necessary to have more even spacing of the data qubit ions.

Next the drive amplitudes of the 355 nm IA Raman beams are calibrated. We calibrate the amplitudes to yield 10 $\mu$s counter-propagating (IA and global) and 25 $\mu$s co-propagating IA $\pi$-times on the $\ket{0}\rightarrow\ket{1}$ transition. In the co-propagating IA configuration, two tones are applied to the AOM for the IA beams while one of those tone amplitude (\toneone/) is swept. For the counter-propagating beam geometry, the tone applied to the IA beams' AOM is swept while the other tone, on the global beam, is kept fixed. We measure Rabi oscillations versus amplitude with a fixed gate duration and determine the saturation amplitude $a_{sat}$ and scaling factor $\Xi$ of the IA beams' AOM in equation \eqref{eq:aom-omega} from a fit to equation \eqref{eq:ampscan-fitfn} (see Section~\ref{sec:arbangle}). Note that in the co- and counter-propagating single-qubit case (unlike the two-qubit case described in Section~\ref{sec:arbangle}), we sweep a single tone applied to the IA beam, not the tone applied to the global. Figure \ref{fig:linenum_ampscan_cal}b and c show the observed Rabi oscillations as the pulse amplitude is swept for the IA beams. For the amplitude scan, the gate duration is increased to 250 $\mu$s (co-propagating, 10x the desired $\pi$ time) or 30 $\mu$s (counter-propagating, 3x the desired $\pi$ time) to generate sufficient oscillations before AOM saturation to ensure a precise fit.  The drive amplitude for the desired $\pi$-time is calculated using equation \eqref{eq:aom-omega} with the fit $a_{sat}$ and $\Xi$. 

The next step in the process is to identify the radial motional modes for both Raman sideband cooling and MS gate calibration. Sideband cooling is applied to all radial motional modes for the pulse amplitude scans described above and the two-qubit gate calibrations described below.

\begin{figure}
    \centering
    \includegraphics[width=0.47\textwidth]{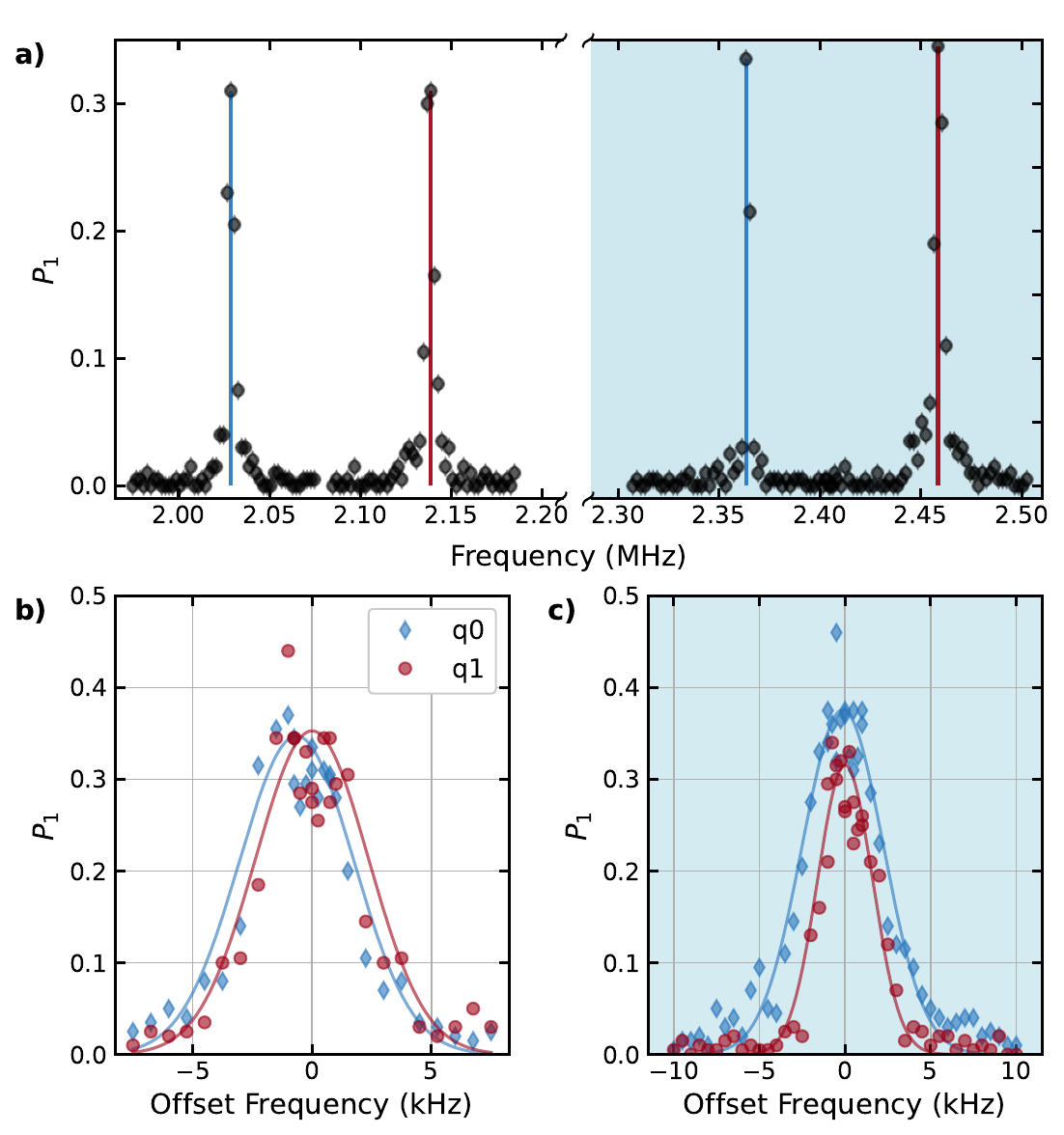}
    \caption{Calibration of radial motional sidebands using qubits $q_0$ and $q_1$. a) Rough calibration using peak maxima. Vertical lines indicate the measured sideband center frequencies. b) and c) Fine scan of lower (b) and upper (c) sidebands centered on peaks found in a). The lower (upper) manifold sidebands are calibrated in parallel, measuring one sideband with each qubit. Solid lines are gaussian fits to the data. Upper sideband data is highlighted with a light blue background. The sidebands in b) and c) correspond to sidebands with the same color vertical line in a).
    }
    \label{fig:side_band_cal}
\end{figure}

Figure \ref{fig:side_band_cal}a shows a coarse frequency sweep of the lower (<2.25MHz) and upper (>2.25 MHz) radial sidebands for two ions. Sideband locations are identified using a peak finding routine. Finer scans, centered around the peaks in Fig. \ref{fig:side_band_cal}a, are shown in Fig. \ref{fig:side_band_cal}b and c for the lower and upper sidebands, respectively. Each sideband in the lower (or upper) manifold is calibrated in parallel by using one ion for each sideband mode. 
For chain lengths of three ions or more, the ion-mode assignment is chosen such that each ion is measuring a mode with which it has a relatively strong coupling. The data is then fit to a parameterized Gaussian function to find sideband center frequencies. The calibrated frequencies are then used for both sideband cooling and MS gate calibrations.

\begin{figure}
    \centering
    \includegraphics[width=0.47\textwidth]{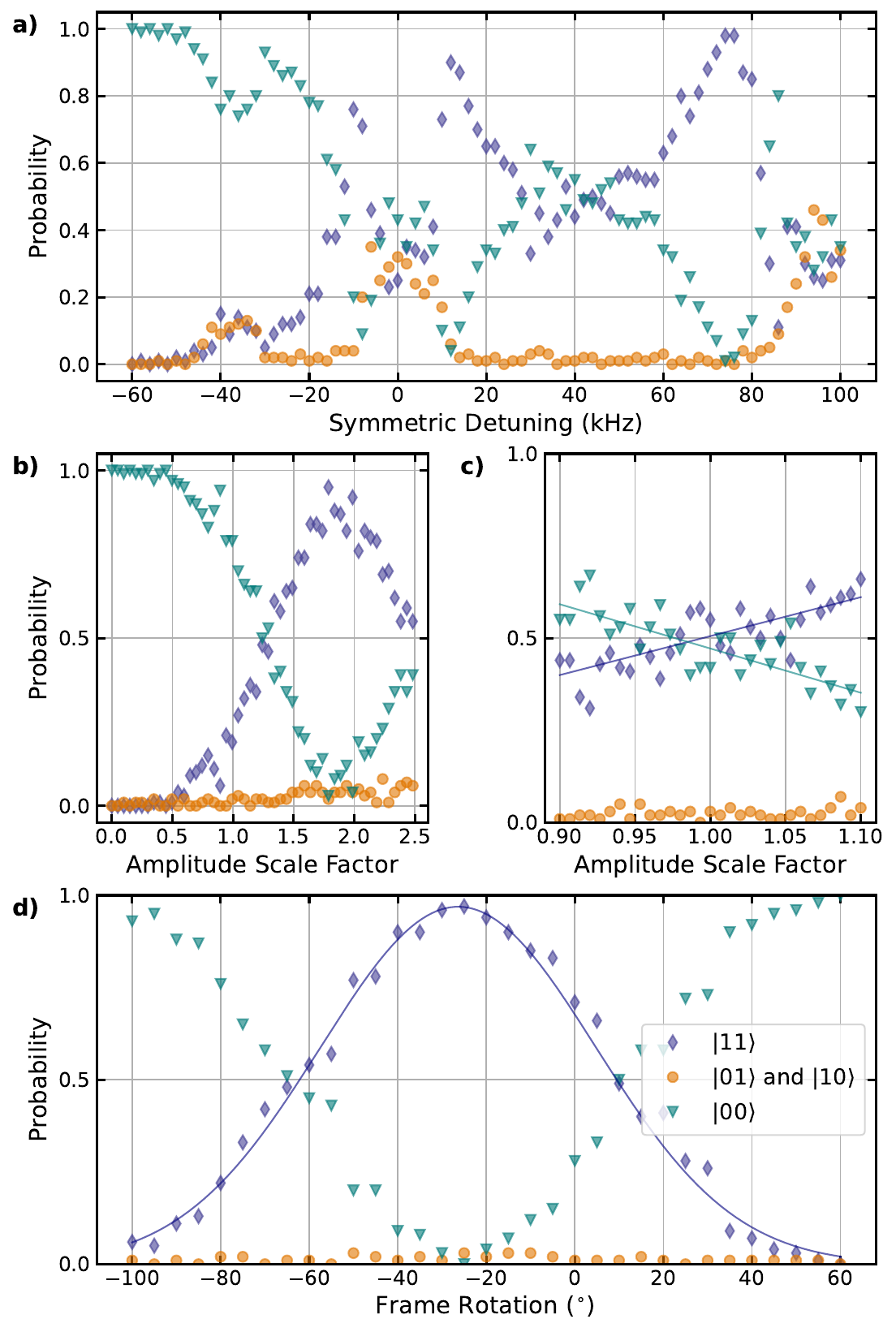}
    \caption{Calibration of the $MS(\theta)$ gate. a) Sweep of symmetric detuning, $\delta_0$, of Raman beams from red and blue sidebands  b) Amplitude scaling is varied coarsely to find the crossing of zero bright and two bright populations. c) Fine scan of amplitude scale factor with linear fits to find crossing of $\ket{00}$ and $\ket{11}$. d) Scan of the frame rotation. $\ket{11}$ is fit to a Gaussian (solid line) to find frame rotation setting.
    }
    \label{fig:ms_gate_cal}
\end{figure}

 Two-qubit MS gates are implemented by applying the global beam in combination with tones on the IA beams' AOM that drive transitions near the radial red and blue motional sidebands of the participating modes. A first indication of entanglement can be realized with a symmetric detuning scan that steps the red and blue tones such that they move symmetrically outward from the carrier transition with increasing detuning. A crossing of the zero and two bright populations with one bright suppressed indicates a detuning that results in entanglement. Figure \ref{fig:ms_gate_cal}a shows a symmetric detuning sweep between two motional modes. The crossings of zero bright and two bright populations near -15 and +40 kHz are possible choices for the MS gate detuning. As described in Section~\ref{sec:freq_robust}, the radial modes used for entangling gates are chosen to give the largest coupling strength (Lamb-Dicke factor) for a particular ion pair and sideband detunings are chosen such that contributions from both modes add constructively. In this case, the region around 40 kHz is the region where contributions from both modes add constructively. In practice, we rarely do this calibration, as we have already measured the sideband frequencies and for given chain lengths already have selected the desired detuning to realize a frequency robust gate. However, it is an instructive measure for understanding the dynamics of the MS gate. Instead, we typically utilize the method described below.

Since we've preselected the detuning to maximize coupling and frequency robustness, we now match the amplitude of the drives to generate the necessary entanglement at that detuning. In practice, this involves finding the ratio $\zeta_{\rm br}$ such that the coherence of the gate is maximized. We utilize the microwave Ramsey-echo sequence described in Fig.~\ref{fig:microwaveecho} to find the appropriate $\zeta_{\rm br}$ for each ion. Once that ratio has been selected per ion, we then sit at the desired detuning per MS gate pair and apply an overall scaling factor, $\kappa$ for the the red and blue amplitudes to generate equal zero and two bright populations, an indication of maximal entanglement i.e. an $MS(\pi/2)$ gate.  Figure \ref{fig:ms_gate_cal}b and c show coarse and fine sweeps of the pulse amplitude scaling that is applied uniformly to both IA tones (note, that the global beam tone is used to then scale down to arbitrary $\theta$). The coarse and fine sweeps are measured sequentially with the scaling updated between scans, hence the scaling of the fine scan very near unity. A linear fit to the to the data in Fig. \ref{fig:ms_gate_cal}c is used to find the final amplitude scaling.

Lastly, we compensate for any residual light shifts by applying a frame rotation as described in Section \ref{sec:lightshifts}. Beginning in the $\ket{00}$ state a sequence of $M \times MS(\pi/M)$ gates of duration $\tau$ is applied to a pair of ions to find the value of the frame rotation $\phi_{f, i}(\tau)$ that maximizes the $\ket{11}$ population. Figure \ref{fig:ms_gate_cal}d shows an example of a frame rotation calibration for a $MS(\pi/2)$ (M=2).  Two MS gates are applied sequentially and the populations are recorded versus the frame rotation angle. The peak of a Gaussian fit specifies the calibrated frame rotation, $\phi_{f,i}(\tau)$ for an $MS(\pi/2)$. The process is repeated for all ion pairs. We also perform the same measurement at M=32, and then linearly interpolate between the two calibrations to determine the necessary $\phi_{f, i}(\tau)$ for any $\theta$.

\begin{figure}
    \centering
    \includegraphics[width=0.47\textwidth]{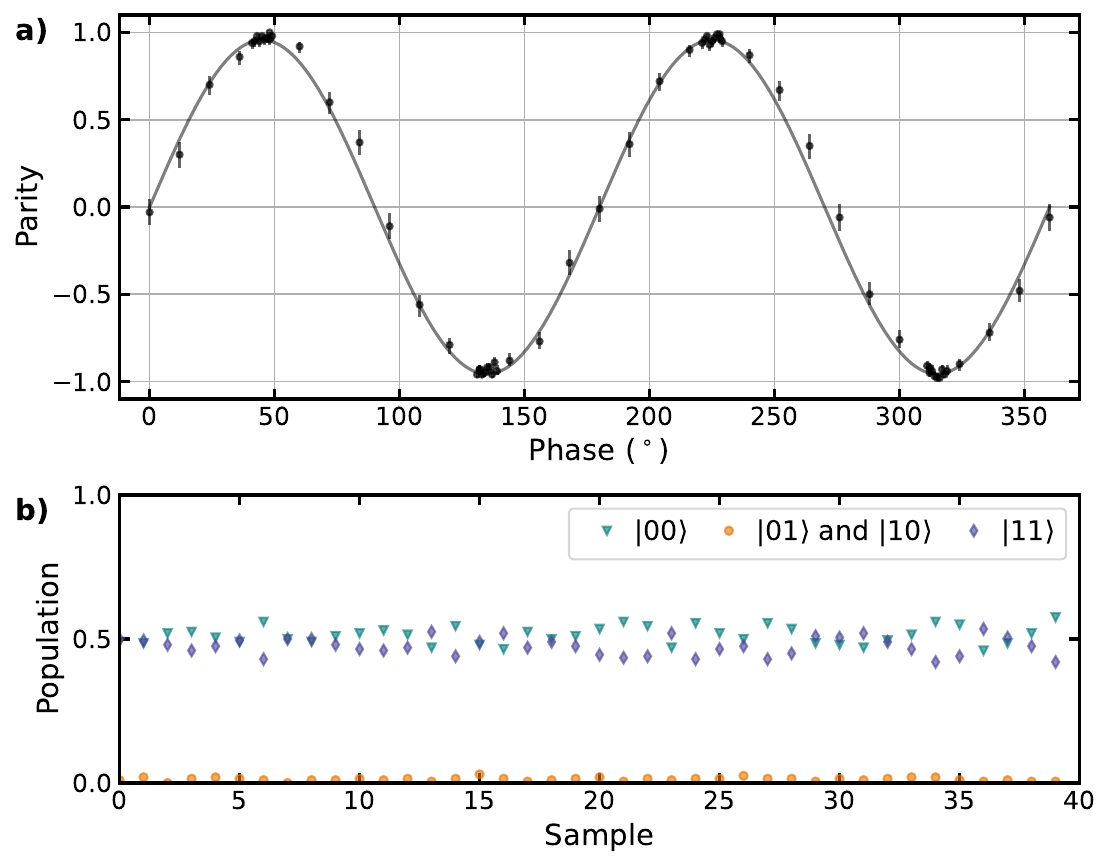}
    \caption{Measurements used to estimate two-qubit gate fidelity. a) Parity scan for calculation of two-qubit gate fidelity. b) State populations following a two-qubit gate.
    }
    \label{fig:parity}
\end{figure}

Two-qubit gate fidelities, as described in Section~\ref{sec:arbangle} are characterized by both a spin probability scan and parity scans. Figure \ref{fig:parity}a a parity measurement with populations in Fig. \ref{fig:parity}b  yielding a fidelity of $0.972^{+0.003}_{-0.004}$ with 95\% confidence intervals on the parity amplitude fit.

\iftoggle{ieeeformat}{
\subsection{Fidelities for Various Chain Lengths}}{
\appsection{Fidelities for Various Chain Lengths}
}
\begin{figure}[h!!!]
    \centering
    \includegraphics[width=0.47\textwidth]{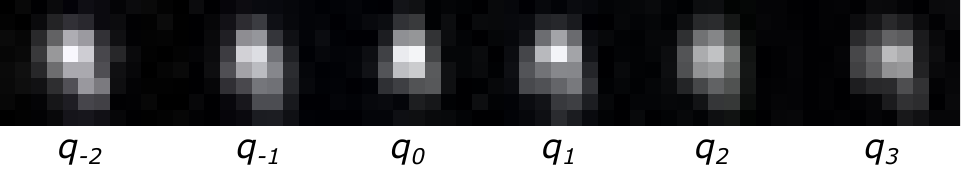}
    \caption{Ions are indexed starting from the center of the
             trap such that increasing chain length does not change the
             relative position of each ion index. For example, $q_0$ is always
             in the center for odd-numbered chains and center-left in
             even-numbered chains.
    }
    \label{fig:ion_labels}
\end{figure}
\label{app:chain_fidelities}
The following tables were compiled using population measurement only fidelity estimations as described in Ref.~\cite{yale2024superstaq, ChowThesis}.
Ion ordering sets the center of the chain (or left-center in an even chain) to be $q_0$, see Fig.~\ref{fig:ion_labels}.
The radial motional sideband mode indexing presented here is different than the mode indexing used in Fig.~\ref{fig:detunings} and the fourth-order light shift discussion. Here, the index starts with the highest frequency mode (the center of mass mode) within the radial manifold of interest and works towards the lowest frequency. For example, in a three ion chain (within one of the two radial manifolds), the mode order would be: [0 = center of mass, 1 = tilt, 2 = zig-zag]. 

\begin{table}[h!!!]
    \centering
    \begin{tabular}{|c|c|c|c|c|c|}
\hline Pair & $q_i$ & $q_j$ & Est. Fidelity ($\sqrt{P_{11}}$) (\%) & Mode & Detuning\\
       index &  &   &   &   index   &(kHz)\\
\hline 0 & 0 & 1 & 99.0\,[+0.1,-0.1] & 1 & 52 \\
\hline
    \end{tabular}
    \caption[Two-ion MS gate fidelity and settings]{M{\o}lmer-S{\o}rensen gate settings and fidelity estimate in a 2 ion chain, as measured by the square root of the $\ket{11}$ population after two stacked gates.} 
    \label{tab:ms-2ion}
\end{table}

\begin{table}[h!!!]
    \centering
    \begin{tabular}{|c|c|c|c|c|c|}
        \hline Pair & $q_i$ & $q_j$ & Est. Fidelity ($\sqrt{P_{11}}$) (\%) & Mode & Detuning\\
               index &  &   &   &   index   &(kHz)\\
        \hline 0 & 0 & 1 & 98.2\,[+0.1,-0.1] & 2 & 30 \\
        \hline 1 & -1 & 0 & 98.2\,[+0.1,-0.1] & 2 & 30 \\
        \hline 2 & -1 & 1 & 97.7\,[+0.1,-0.2] & 2 & 42 \\
        \hline
    \end{tabular}
    \caption[All MS gate pairs for a three ion chain]{M{\o}lmer-S{\o}rensen gate fidelity estimate for all pairs of ions in a 3 ion chain, as measured by the square root of the $\ket{11}$ population after two stacked gates. Note that gates performed on the center ion ($q_{0}$) are not balanced gates as the center ion does not participate in antisymmetric motional modes.} 
    \label{tab:ms-3ion}
\end{table}

\begin{table}[h!!!]
    \centering
    \begin{tabular}{|c|c|c|c|c|c|}
    \hline Pair & $q_i$ & $q_j$ & Est. Fidelity ($\sqrt{P_{11}}$) (\%) & Mode & Detuning\\
       index &  &   &   &   index   &(kHz)\\
    \hline 0 & 0 & 1 & 98.8\,[+0.2,-0.3] & 3 & 52 \\
    \hline 1 & 0 & -1 & 98.4\,[+0.3,-0.3] & 2 & 30 \\
    \hline 2 & 1 & -1 & 97.9\,[+0.3,-0.3] & 3 & 34 \\
    \hline 3 & 0 & 2 & 97.7\,[+0.3,-0.4] & 3 & 36 \\
    \hline 4 & 1 & 2 & 98.5\,[+0.2,-0.3] & 2 & 30 \\
    \hline 5 & -1 & 2 & 98.0\,[+0.3,-0.3] & 2 & 30 \\
\hline
\end{tabular}
\caption{M{\o}lmer-S{\o}rensen gate fidelity estimate for all pairs of ions in a 4 ion chain, as measured by the square root of the $\ket{11}$ population after two stacked gates.} 
\end{table}

\begin{table}[h!!!]
    \centering
    \begin{tabular}{|c|c|c|c|c|c|}
        \hline Pair & $q_i$ & $q_j$ & Est. Fidelity ($\sqrt{P_{11}}$) (\%) & Mode & Detuning\\
       index &  &   &   &   index   &(kHz)\\
        \hline 0 & 0 & 1 & 98.0\,[+0.3,-0.3] & 4 &  30 \\
        \hline 1 & 0 & -1 & 97.8\,[+0.3,-0.3] & 4 & 30 \\
        \hline 2 & 1 & -1 & 97.7\,[+0.3,-0.4] & 4 & 40 \\
        \hline 3 & 0 & 2 & 98.0\,[+0.3,-0.3] & 2 & 22 \\
        \hline 4 & 1 & 2 & 98.0\,[+0.3,-0.3] & 2 & 20 \\
        \hline 5 & -1 & 2 & 96.8\,[+0.4,-0.4] & 3 & 20 \\
        \hline 6 & 0 & -2 & 97.9\,[+0.3,-0.3] & 2 & 20 \\
        \hline 7 & 1 & -2 & 96.3\,[+0.4,-0.4] & 3 & 18 \\
        \hline 8 & -1 & -2 & 97.9\,[+0.3,-0.3] & 3 & 20 \\
        \hline 9 & 2 & -2 & 97.2\,[+0.3,-0.4] & 2 & 25 \\
        \hline
    \end{tabular}
    \caption[All MS gate pairs for a five ion chain]{M{\o}lmer-S{\o}rensen gate fidelity for all pairs of ions in a 5 ion chain, as measured by the square root of the $\ket{11}$ population after two stacked gates. Note that gates performed on the center ion ($q_{0}$) are not balanced gates as the center ion does not participate in antisymmetric motional modes.} 
    \label{tab:ms-5ion}
\end{table}

\begin{table}[h!!!]
    \centering
    \begin{tabular}{|c|c|c|c|c|c|}
    \hline Pair  & $q_i$ & $q_j$ & Est. Fidelity ($\sqrt{P_{11}}$) (\%) &Mode   &  Detuning  \\
           index &       &       & [+1\%, -2\%] & index &   (kHz)     \\
    \hline 0 & 0 & 1   & 97.1 & 5 & 43 \\
    \hline 1 & 0 & -1  & 95.6 & 4 & 30 \\
    \hline 2 & 1 & -1  & 96.8 & 5 & 35 \\
    \hline 3 & 0 & 2   & 95.4 & 5 & 36 \\
    \hline 4 & 1 & 2   & 97.8 & 4 & 35 \\
    \hline 5 & -1 & 2  & 93.5 & 4 & 30 \\
    \hline 6 & 0 & -2  & 95.9 & 2 & 22 \\
    \hline 7 & 1 & -2  & 95.7 & 3 & 20 \\
    \hline 8 & -1 & -2 & 95.9 & 3 & 26 \\
    \hline 9 & 2 & -2  & 95.4 & 3 & 26 \\
    \hline 10 & 0 & 3  & 94.0 & 3 & 23 \\
    \hline 11 & 1 & 3  & 94.7 & 2 & 22 \\
    \hline 12 & -1 & 3 & 94.7 & 3 & 25 \\
    \hline 13 & 2 & 3  & 93.6 & 2 & 20 \\
    \hline 14 & -2 & 3 & 95.2 & 2 & 20 \\
    \hline
    \end{tabular}
    \caption[All MS gate pairs for a six ion chain]{M{\o}lmer-S{\o}rensen gate fidelity for all pairs of ions in a 6 ion chain, as measured by the square root of the $\ket{11}$ population after two stacked gates.} 
    \label{tab:ms-ion}
\end{table}

\newpage
\bibliography{refs} 
\iftoggle{ieeeformat}{
\bibliographystyle{IEEEtran}
}{}
\iftoggle{ieeeformat}{
\EOD}{}
\end{document}